\documentclass[11pt,a4paper]{article}
\usepackage{jheppub}
\usepackage{verbatim}
\usepackage{mciteplus}
\usepackage{hyperref}
\usepackage{amssymb}
\usepackage{amsthm}
\usepackage{amstext}
\usepackage{amsmath}
\usepackage{amsfonts}
\usepackage{dsfont}
\usepackage{cancel}
\usepackage{jheppub}
\usepackage{bbm}
\usepackage{wrapfig}
\usepackage{graphicx}
\usepackage{subfigure}
\usepackage{pstricks}
\usepackage{color}

\usepackage{enumitem}

\usepackage{xparse}

\def\Tr{\mathop{\rm Tr}}

\def\eps{\epsilon}

\newcommand{\la}{\langle}
\newcommand{\ra}{\rangle}

\NewDocumentCommand\Tensor{mmggg}{
#1 \otimes #2 \IfNoValueTF{#3}{}{\otimes #3} \IfNoValueTF{#4}{}{\otimes #4} \IfNoValueTF{#5}{}{\otimes #5}
}

\theoremstyle{plain}

\numberwithin{equation}{section}

\title{A first look at the function space for planar two-loop six-particle Feynman integrals}
\author[a]{Johannes Henn}
\author[b]{Tiziano Peraro}
\author[c]{Yingxuan Xu}
\author[d,e,a]{and Yang Zhang}

\preprint{MPP-2021-200, USTC-ICTS/PCFT-21-36, HU-EP-21/57-RTG}

\affiliation[a]{Max-Planck-Institut f\"ur Physik, Werner-Heisenberg-Institut, D-80805 M\"unchen, Germany}
\affiliation[b]{ Dipartimento di Fisica e Astronomia, Universit\`a di Bologna e INFN, Sezione di Bologna, via Irnerio 46, I-40126 Bologna, Italy}
\affiliation[c]{Humboldt-Universit\"at zu Berlin, Institut f\"ur
  Physik, Newtonstra\ss e 15, 12489 Berlin, Germany}
\affiliation[d]{Interdisciplinary Center for Theoretical Study, University of Science and Technology of China,
Hefei, Anhui 230026, China}
\affiliation[e]{Peng Huanwu Center for Fundamental Theory, Hefei, Anhui 230026, China}

\emailAdd{henn@mpp.mpg.de}
\emailAdd{tiziano.peraro@unibo.it}
\emailAdd{yingxu@physik.hu-berlin.de}
\emailAdd{yzhphy@ustc.edu.cn}

\abstract{Two-loop corrections to scattering amplitudes are crucial
  theoretical input for collider physics. 
Recent years have seen tremendous advances in computing Feynman integrals, scattering amplitudes, and cross sections for five-particle processes.
In this paper, we initiate the study of the function space for planar two-loop six-particle processes.
 We study all genuine six-particle Feynman integrals, and derive the differential equations they satisfy on maximal cuts.
Performing a leading singularity analysis in momentum space, and in Baikov representation, we find an integral basis that puts the differential equations into canonical form. 
 The corresponding differential
  equation in the eight independent kinematic variables is derived with the finite-field reconstruction method and the
  symbol letters are identified. We identify the dual conformally invariant hexagon alphabet known from maximally supersymmetric Yang-Mills theory as a subset of our alphabet. This paper constitutes an important step in the analytic calculation of planar two-loop six-particle Feynman integrals.
 }


\begin{document}

\maketitle

\section{Introduction}
In the coming years, the future high-luminosity Large Hadron Collider
(LHC) and proposed Circular Electron Positron Collider (CEPC), Future Circular Collider (FCC) and International Linear Collider (ILC) experiments will accumulate a large
amount of data, which need to be compared by precision theoretical predictions
for particle scattering cross sections~\cite{Amoroso:2020lgh, Proceedings:2018jsb, Heinrich:2020ybq, Heinrich:2017una, Mangano:2017tke, Azzi:2019yne,CidVidal:2018eel}. To get the relevant cross
sections at NNLO, one crucial task is the evaluation of multi-loop
Feynman integrals with several external legs. 

Recent years have seen tremendous progress in computing five-particle Feynman integrals, scattering amplitudes at two loops, and even complete cross sections.
All the massless two-loop five-point integrals were calculated analytically in a series of papers~\cite{Gehrmann:2015bfy,Chicherin:2017dob,Abreu:2018rcw,Abreu:2018aqd,Chicherin:2018mue,Chicherin:2018yne,Chicherin:2018old}. 
The result provides a complete set of two-loop Feynman integrals for any massless $2\rightarrow3$ scattering process, which can be used for three-jet production at hadron colliders to NNLO in QCD.
Very importantly, a dedicated computer implementation for evaluating the Feynman integrals in the physical region in an efficient and reliable way was provided in reference \cite{Chicherin:2020oor}.
First applications of the analytic progress include three-photon production \cite{Chawdhry:2020for,Abreu:2020cwb,Kallweit:2020gcp}, three-jet production \cite{Abreu:2021oya,Czakon:2021mjy}, and diphoton plus jet production \cite{Chawdhry:2021mkw,Chawdhry:2021hkp,Agarwal:2021vdh,Badger:2021imn,Badger:2021ohm}.

Also very recently, first results on two-loop five-point integrals with one off-shell leg have become available \cite{Abreu:2020jxa,Canko:2020ylt,Abreu:2021smk}.
These integrals are important ingredients for two-loop scattering amplitudes for two-jet-associated W-boson/Z-boson production in QCD, see \cite{Badger:2021nhg}. 

From the above paragraphs it is clear that the state of the art in analytically computing two-loop Feynman integrals is five particles. 
Moreover, we see that when new results for Feynman integrals become available, this constitutes a game changer and paves the way towards obtaining scattering amplitudes, and eventually full cross sections.
This motivates us to aim at the two-loop six-point massless integrals, where very little is known in general.
The results for five-point integrals with one off-shell leg give a subset of the integrals needed for genuine six-particle scattering. 
Moreover, we can get some information (and motivation) from ${\mathcal N}=4$ super Yang-Mills (sYM), where the function space for planar six-particle scattering is believed to be known to all loop orders, and is closely linked to cluster algebras \cite{Goncharov:2010jf}. The relevant function space is beautifully given by nine alphabet letters that define a certain class of iterated integrals.
Remarkably, the knowledge of the function space, together with physical properties of amplitudes, makes it possible to bootstrap results to very high loop orders, as reviewed in \cite{Caron-Huot:2020bkp,Henn:2020omi}. 
Unfortunately, this result is closely linked to additional symmetries of planar ${\mathcal N}=4$ sYM that are broken in QCD, and as a result, the corresponding integrals in QCD are considerably more complicated.

Already for five massless particles, the scattering kinematics, consisting of five independent dimensionful Mandelstam variables turned out to be one of the main challenges in QCD. One of the key insights was the discovery that those variables enter the Feynman integrals in a very precise way. More precisely, it turned out that all relevant Feynman integrals can be described by a function space with certain symbol letters, and there are exactly $31$ symbol letters (that each depends in a precise way on the five-particle kinematics), as initially conjectured in \cite{Chicherin:2017dob}.

 At six massless particles, there are eight independent dimensionful kinematic variables, so that one can imagine a very complicated function space. One of our goals will therefore be to identify appropriate symbol letters that describe the function space, generalizing the nine dual conformal letters from ${\mathcal N}=4$ sYM, and the letters known from five-particle integrals with one off-shell leg. 
Our second goal will be to initiate the systematic calculation of all planar two-loop six-particle integrals. The reason for this is that, given the smaller amount of symmetries that QCD enjoys compared to ${\mathcal{N}=4}$ sYM, it is likely that in practice the knowledge of the symbol alphabet, while important, is not sufficient to bootstrap full QCD amplitudes.
Our method of choice to achieve these two goals is the canonical differential equation method \cite{Henn:2013pwa,Henn:2014qga}, that was instrumental in obtaining the five-particle results mentioned above.
Let us quickly review the main steps and challenges. 

A key tool in these computations is the integration-by-parts (IBP) identities~\cite{Tkachov:1981wb,Chetyrkin:1981qh} that arise from the vanishing integration of total derivatives. 
In practice, the IBP identities generate linear relations between loop integrals with a specific set of propagators but with different numerators, allowing any integral to be expressed in terms of a finite basis of integrals \cite{Smirnov:2010hn}, so-called master integrals. In this way, the problem is reduced to that of evaluating the master integrals. 

A very successful method for evaluating the master integrals is the differential equation with respect to kinematic invariants~\cite{Kotikov:1990kg,
Remiddi:1997ny}. For each set of Feynman integrals, one obtains a set of first-order differential equations. These equations, together with an appropriate boundary condition, determine the answer in principle. However, in general this solution is very hard to obtain if the differential equation matrix is complicated.
One of the keys to the solution of the differential
equation is to choose an optimal basis of integrals that leads to a
system of differential equations in a canonical form
\cite{Henn:2013pwa,Henn:2014qga}, in which the loop integrals have uniform
degree of transcendentality (UT), also called transcendental weight.

A key insight, originally observed for integrals appearing in maximally supersymmetric Yang-Mills theory \cite{ArkaniHamed:2010gh}, is that the UT property of the integrated functions can be understood from properties of the rational loop {\it integrand}. It was observed that the UT integrals in that have integrands that can be written in a so-called {\it dlog} form \cite{Arkani-Hamed:2014via}, that is, they only have single poles in the integration variables. Moreover, their maximal residues, also called leading singularities, are normalized to kinematic-independent constants. These properties have hence been used  for finding UT integrals beyond ${\mathcal{N}}=4$ sYM (see e.g. \cite{Wasser:2018qvj,Henn:2020lye}, which includes an algorithmic implementation.) In particular they have been crucial for calculating the two-loop five-particle integrals mentioned above. 

In the most basic version, the integrand analysis is done in four dimensions (or in another integer dimension). Quite remarkably, for most cases this is sufficient to predict the UT property of the integrals in dimensional regularization, with $D=4-2\eps$, for any order in $\eps$. However, especially when many kinematic variables are involved, it was noticed that a more refined analysis is necessary, that also includes aspects beyond four dimensions. This is not too surprising, if one considers that it is possible to write down evanescent integrands, i.e. integrands that vanish when the loop momentum is taken to be four-dimensional, but which give non-zero answers after integration. In the case of six particles, information on four-dimensional integrands is available from \cite{Bourjaily:2019iqr,Bourjaily:2021hcp}. Moreover, the integrand of two-loop six-gluon all-plus scattering amplitudes is available in analytic form in \cite{Badger:2016ozq}.

A very useful way of including more than the four-dimensional information is to perform the leading singularity analysis in Baikov parametrization. This idea has been used for the finding of two-loop five-point massless UT integrals \cite{Chicherin:2018old} as well as two-loop double box UT integrals with four external masses \cite{Dlapa:2021qsl}. (See also \cite{Chen:2020uyk} which discusses this method in the context of intersection theory~\cite{Mastrolia:2018uzb,Frellesvig:2019kgj,Frellesvig:2019uqt,Frellesvig:2020qot}.)

Already at five particles, this program faces enormous practical challenges, in terms of the necessary computer algebra and running time, the size of intermediate and final expressions. For this reason our analysis of the uniform weight properties of integrals is crucial, as the differential equations are known to  simplify dramatically. That being said, already in the case of five-particle amplitudes with one off-shell leg, they involve numerous different singularities, depending on six kinematic variables. Each singularity describes a physically interesting kinematic configuration, such as collinear limits, thresholds, and so on.
The general six-particle case we set out to study involves nine kinematic variables, and one anticipates even more singularities, which makes it particularly challenging.

In view of the technical challenges, obtaining the full differential equations in one step is a difficult task. 
However, we can profit from the fact that the differential equation method can be combined naturally with (generalized) unitarity cuts. 
This fact is used for example in the reverse unitarity method \cite{Anastasiou:2002yz}. 
What is important in the present context is that cut integrals, i.e. integrals where certain propagators are replaced by delta functions, satisfy the same differential equations, albeit with different boundary conditions.
Moreover, as only integrals with certain propagators present can contribute to a given cut, this allows us to focus on a given integral sector at a given time.
Each integral sector (corresponding to a set of propagators) corresponds to a block on the diagonal of the differential equations matrix.
In this paper we analyze the planar two-loop six-point integrals (beyond the ones that correspond to two-loop five-point integrals with one
massive external leg) on maximal cuts. 

We use a variant of dimensional regularization called t'Hooft-Veltman (tHV) scheme, where external states are four-dimensional while internal (loop) states are $D$-dimensional, with $D=4-2 \epsilon$.  
This differs from the so-called conventional dimensional regularization (CDR) where everything is in $D$ dimensions.  This is the same scheme commonly used for helicity amplitudes, including all recent five-point results. 
However, up to five external legs, the IBPs and the master integrals are the same in the two schemes, since there are at most four independent external momenta (amplitudes can differ because there you also have polarization states).  Six external legs is the lowest multiplicity where one sees a scheme difference also in the calculation of the reduction and the master integrals.  A recent review on this topic is \cite{Gnendiger:2017pys}.

We perform a leading singularity analysis in Baikov representation in order to search for UT integrals with complicated kinematics \cite{Chicherin:2018old,Dlapa:2021qsl,Chen:2020uyk}.
We then use finite-field reconstruction for the IBP reduction \cite{vonManteuffel:2014ixa} and for deriving
 the differential equation. The diagrams considered in this paper have
  complicated kinematics and the traditional IBP reduction is
  difficult. To overcome this, we use the cutting edge finite-field reconstruction
  approach with the package {\sc FiniteFlow} \cite{Peraro:2016wsq,Peraro:2019svx, Peraro:2019okx} for the
  computations. In doing so, we find momentum twistor parameterization \cite{Hodges:2009hk, Mason:2009qx} useful, as it
  rationalizes certain Gram-determinant square roots.  
  
  In this way, we are able to reconstruct the differential equations on the maximal cuts. We verify that they take a UT form for the integrals we identified with the Baikov analysis.
Furthermore, we identify and present the symbol letters appearing in the canonical differential equations. 
This constitutes a crucial step in the calculation of the planar six-particle two-loop integrals, and in identifying the relevant function space. We anticipate that the latter will prove useful for a better analytic understanding of six-particle scattering processes, for instance in the context of bootstrap methods.

This paper is organized as follows: In section 2, we introduce our conventions for 
six-particle kinematics, including a useful momentum twistor parametrization. We also discuss the two-loop six-particle integrals and define the integral sectors analyzed in this paper. In section 3 we discuss the main methods used: after outlining the canonical differential equation method, we briefly review the integrand analysis for finding uniform weight integrals, and discuss how we employ finite field methods to construct the canonical differential equations. Sections 4 and 5 contain our main results: In section 4, we give, for each integral sector, a set of master integrals that lead to canonical differential equations on the cut, and we comment on the alphabet letters appearing in the equations. Section 5 summarizes our results for the function alphabet. These two sections are complemented with ancillary files that contain our results in computer-readable form. 
Finally, in section 6, we summarize our results, and discuss future directions.

\section{Conventions and definitions}

\subsection{Planar massless six-particle kinematics}
\label{sec:kinematics}

In this section, we introduce the conventions of the two-loop six-point massless kinematics.

The six external momenta are named as $p_i$'s, $i=1,\ldots 6$ with $p_i^2=0$
and
\begin{equation}
\sum_{i=1}^6 p_i=0\,.
\end{equation}
We recall that throughout this paper, all external momenta are treated as four-dimensional.
The Mandelstam variables are 
\begin{equation}
  \label{eq:2}
  s_{ij}=(p_i+p_j)^2\,,\quad s_{ijk} = (p_i + p_j + p_k)^2 \,,\quad 1\leq i,j,k\leq 6\,.
\end{equation}
Moreover, we define a notation for the Gram determinant 
\begin{equation}
  \label{eq:3}
  G\left(
\begin{array}{ccc}
u_1 &\ldots & u_n\\
v_1 & \ldots & v_n
\end{array}\right) = {\rm det} (u_i \cdot v_j)\,,
\end{equation}
where the R.H.S. is the determinant of the  $n\times n$ matrix with the entries $(u_i
\cdot v_j)$, $1\leq i,j\leq n$. 
We use the abbreviation 
\begin{equation}
  \label{eq:24}
  G(i_1 ,\ldots, i_k) \equiv G\left(
\begin{array}{ccc}
p_{i_1} &\ldots & p_{i_k}\\
p_{i_1} & \ldots & p_{i_k}
\end{array}\right)
,\quad 1\leq i_1, ... i_k \leq 6
\end{equation}

For $D$ dimensional external momenta, there are $9$ independent Mandelstam variables, see e.g. \cite{eden2002analytic}.
In four dimensions, however, five vectors are linearly dependent. The resulting constraint can be written as
\begin{equation}
  \label{eq:25}
  G(1,2,3,4,5)=0\,.
\end{equation}
This means that there are only $8$ independent Mandelstam variables.

In addition to the scalar invariants discussed above, there are also pseudo scalars,
which we write as
\begin{equation}
  \label{eq:11}
  \epsilon_{ijkl}\equiv 4 \sqrt{-1} \ \varepsilon_{\mu_1 \mu_2 \mu_3
    \mu_4} p_i^{\mu_1} p_j^{\mu_2}p_k^{\mu_3}p_l^{\mu_4},\quad 1\leq
  i,j,k,l \leq 6
\end{equation}
where $\varepsilon$ is the 4D Levi-Civita symbol. In addition to the independent Mandelstam variables, one needs one pseudo scalar to fully specify the kinematics \cite{eden2002analytic}.
The latter can be any of the ones appearing in eq. (\ref{eq:11}), for example $ \epsilon_{1234}$.

The pseudo scalars are related to the the Gram determinant $G(i,j,k,l)$ as follows  
\begin{equation}
  \label{eq:26}
  G(i,j,k,l)= \frac{1}{16} \epsilon_{ijkl}^2\,.
\end{equation}
From this one sees that the new information contained in $\eps_{ijkl}$ is a sign, $\eps_{ijkl} = \pm 4 \sqrt{G(i,j,k,l)}$.
Since $\eps_{1234} \eps_{ijkl}$ can be expressed in terms of Mandelstam variables, only the sign for one choice of $\{i,j,k,l\}$ is needed \cite{eden2002analytic}.

Other useful kinematic quantities that we will employ in this paper are,
\begin{align}
  \label{trace_pm}
  {\rm Tr}_{+}(ijkl)&\equiv  s_{ij}s_{kl} -s_{ik}s_{jl} +s_{il}s_{jk}
  +\epsilon_{ijkl} \,, \\
 {\rm Tr}_{-}(ijkl)&\equiv s_{ij}s_{kl} -s_{ik}s_{jl} +s_{il}s_{jk}
  -\epsilon_{ijkl}\,.
\end{align}

Frequenly, we also use spinor helicity notations in this paper:
\begin{equation}
  \label{eq:35}
  p_{i\mu} \sigma^\mu_{\alpha \dot \beta}=\lambda_{i\alpha}\tilde
  \lambda_{j\dot \beta} 
\end{equation}
with the Pauli matrices $\sigma^\mu=(I_{2\times 2}, \sigma^1, \sigma^2,
\sigma^3)$. The spinor products are defined as,
\begin{eqnarray}
  \la ij \ra &\equiv \lambda_i ^\alpha \lambda_{j,\alpha }\\ 
  \ [ij]   &\equiv \tilde \lambda_{i ,\dot{\alpha}} \tilde
\lambda_j^{\dot{\alpha }} \,.
\end{eqnarray}
With spinor products, kinematic quantities like \eqref{trace_pm}
can be alternatively expressed as,
\begin{align}
  \label{eq:37}
  {\rm Tr}_{+}(ijkl)& =2 [ij]\la jk \ra [kl] \la li\ra\\
{\rm Tr}_{-}(ijkl)&=2 \la ij \ra [jk] \la kl\ra [li] \,.
\end{align}

\subsection*{Momentum twistor parametrization}
Recall that we treat the external states as four-dimensional. 
In this case one can employ momentum twistor variables \cite{Hodges:2009hk},
that have proven very useful in many contexts of studying scattering amplitudes.
In particular, since momentum twistors are unconstrained (they automatically
solve momentum conservation and massless on-shell conditions), it is
easy to derive identities between functions expressed in them.

Nevertheless, we will often use the variables introduced above in order to denote
final expressions (using, sometimes, an over-complete set of Mandelstam
variables  and pseudo scalars), as we find that this leads to shorter expressions.
To derive these compact expressions, momentum twistors are very useful.

We refer interested readers to \cite{Hodges:2009hk} for more details and definitions of the
momentum twistors. For the purposes of this paper, however, it suffices to say
that we use a particular momentum twistor parametrization, similar to \cite{Badger:2013gxa},
for our kinematic computations.

The latter is
\begin{equation}
  \label{eq:t3}
  Z=\left(
    \begin{array}{cccccc }
    1  & 0  & \frac{1}{x_1}& \frac{1}{x_1}+\frac{1}{x_1 x_2} &
                                                               \frac{1}{x_1}+\frac{1}{x_1
                                                               x_2}+\frac{1}{x_1
                                                               x_2
                                                               x_3} & \frac{1}{x_1}+\frac{1}{x_1
                                                               x_2}+\frac{1}{x_1
                                                               x_2
                                                                      x_3}
                                                                      +
      \frac{1}{x_1
                                                               x_2 x_3
                                                                      x_4}\\
     0 & 1 & 1 & 1 & 1 & 1\\
  0 & 0 & 0 & \frac{x_5}{x_2} & x_6 & 1\\
  0 & 0 & 1 & 1 & x_7 & 1-\frac{x_8}{x_5}
    \end{array}
\right)\, ,
\end{equation}
where $x_1,\ldots, x_8$ are free variables.  With this choice, all $s_{ij}$ and $\epsilon_{ijkl}$ are rational functions
of $x_i$'s, $i=1,\ldots 8$.  The parameterization in \eqref{eq:t3} is
equivalent to the following explicit parameterization,
\begin{align}
  \label{eq:35}
  s_{12}&=x_1 \\
  s_{23}&=x_1 x_5  \\
  s_{13}&= -x_1 \left(x_5-x_8+1\right)  \\
s_{16}& =-\frac{x_1 x_2 x_3 x_4 \left(x_5 x_6-x_8 x_6-x_5
        x_7\right)}{x_5}\\
s_{24}&=\frac{x_1 \left(x_2+1\right) \left(-x_3 x_5-x_5+x_2 x_3
        x_6\right)}{x_2}\\
s_{34}&=-\frac{x_1 \left(-x_2 x_3 x_5-x_3 x_5+x_2 x_3 x_7 x_5-x_5+x_2 x_3 x_6\right)}{x_2}\\
\text{Tr}_+(1234)&= 2 x_1^2 \left(-x_2 x_3 x_5-x_3 x_5+x_2 x_3 x_7
                   x_5-x_5+x_2 x_3 x_6\right)\\
\text{Tr}_+(1236)&=2 x_1^2 x_2 x_3 x_4 \left(x_7
                   x_5-x_5+x_6+x_8-1\right).
\label{momentum_parametrization}
\end{align}
By these formulae, other kinematic variables, $s_{ij}$, $\epsilon_{ijkl}$, $\text{Tr}_\pm(ijkl)$,
are uniquely determined as rational functions in $x_i$'s. The
explicit expression of all kinematic variables in this
parameterization is given in the ancillary file.

The inverse map from the momentum-twistor variables to ``normal''
variables is
\begin{align}
  \label{eq:6}
  x_1 &= s_{12} \\
  x_2&=-\frac{{\rm Tr}_{+}(1234) }{2 s_{12} s_{34}} \\
 x_3&=-\frac{{\rm Tr}_{+}(1345) }{2 s_{45} s_{13}} \\
x_4&=-\frac{{\rm Tr}_{+}(1456) }{2 s_{56} s_{14}} \\
x_5&=\frac{s_{23}}{s_{12}} \\
x_6&=-\frac{{\rm Tr}_{+} (1532)+{\rm Tr}_{+} (1542)}{2 s_{15} s_{12}}\\
x_7&=1+\frac{{\rm Tr}_{+} (1542)+{\rm Tr}_{+} (1543)}{2 s_{15} s_{23}}\\
x_8&=\frac{s_{123}}{s_{12}}
\end{align}
These expressions are simple, and it is easy to rewrite a rational function in
$x_i$'s as a function of $s_{ij}$ and $\epsilon_{ijkl}$. However, it
is not straightforward to directly simplify an expression with $s_{ij}$ and $\epsilon_{ijkl}$.

To repeat, since the $x_i$ are unconstrained variables, they are suitable for automating the computation of $6$-point kinematics with common computer algebra systems.

\subsection{Planar two-loop six-particle integrals}
\label{sec:integral-families}

The goal of this paper is to derive differential equations for planar two-loop six-particle Feynman integrals.
Since the UT integrals for the planar two-loop five-point diagrams
with one massive external leg are known~\cite{Abreu:2020jxa,Canko:2020ylt}, in this paper we focus on the ``genuine'' two-loop six-point planar diagrams. Here ``genuine'' means the
corresponding diagrams which (1) cannot be factorized to two one-loop
diagrams, and  (2) do not have two or more external legs combined.

\begin{figure}[t]
\centering
  \subfigure [Double-pentagon] {\includegraphics[width=5cm]{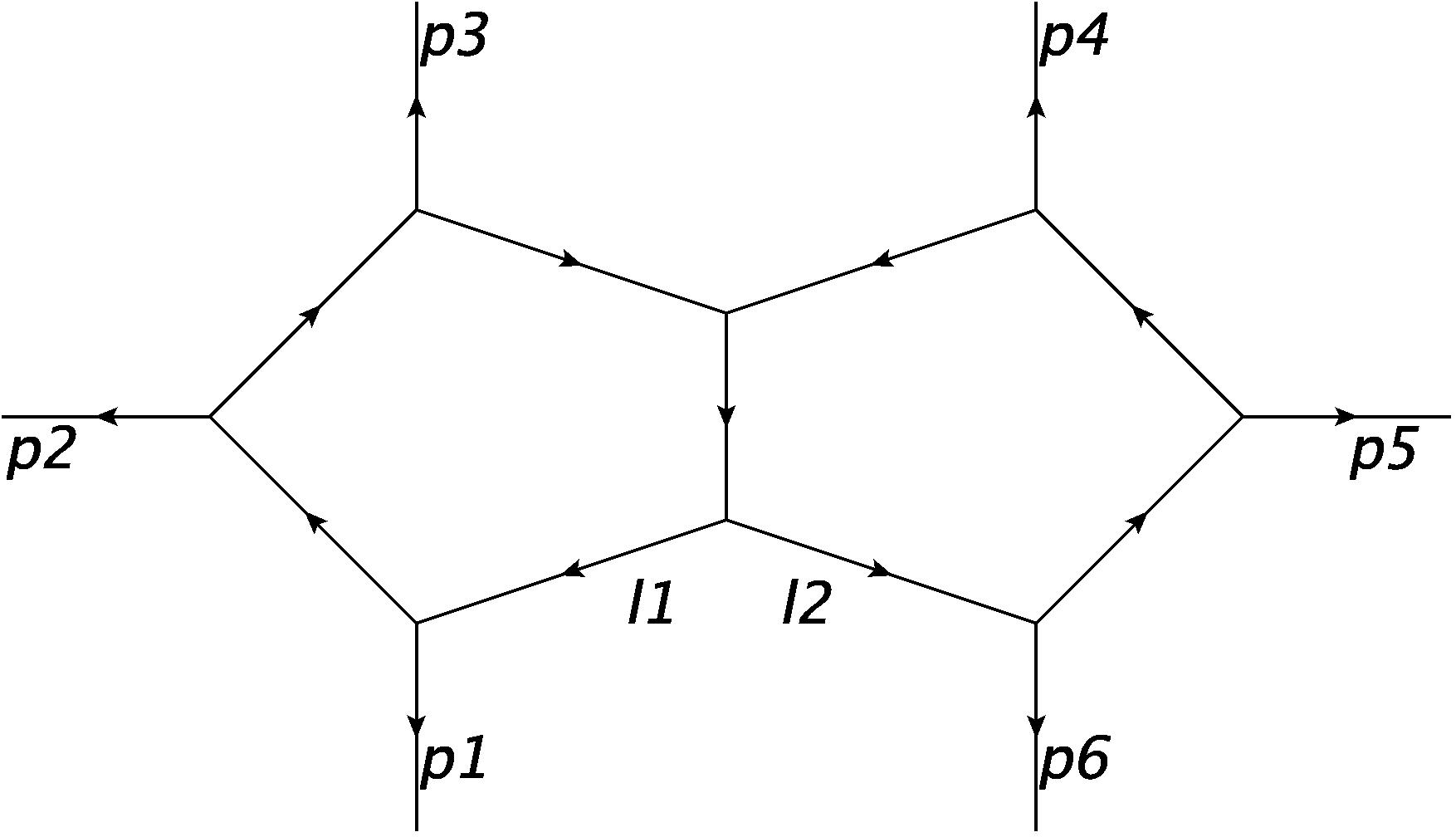}}
  \subfigure [Pentagon-box] {\includegraphics[width=4.5cm]{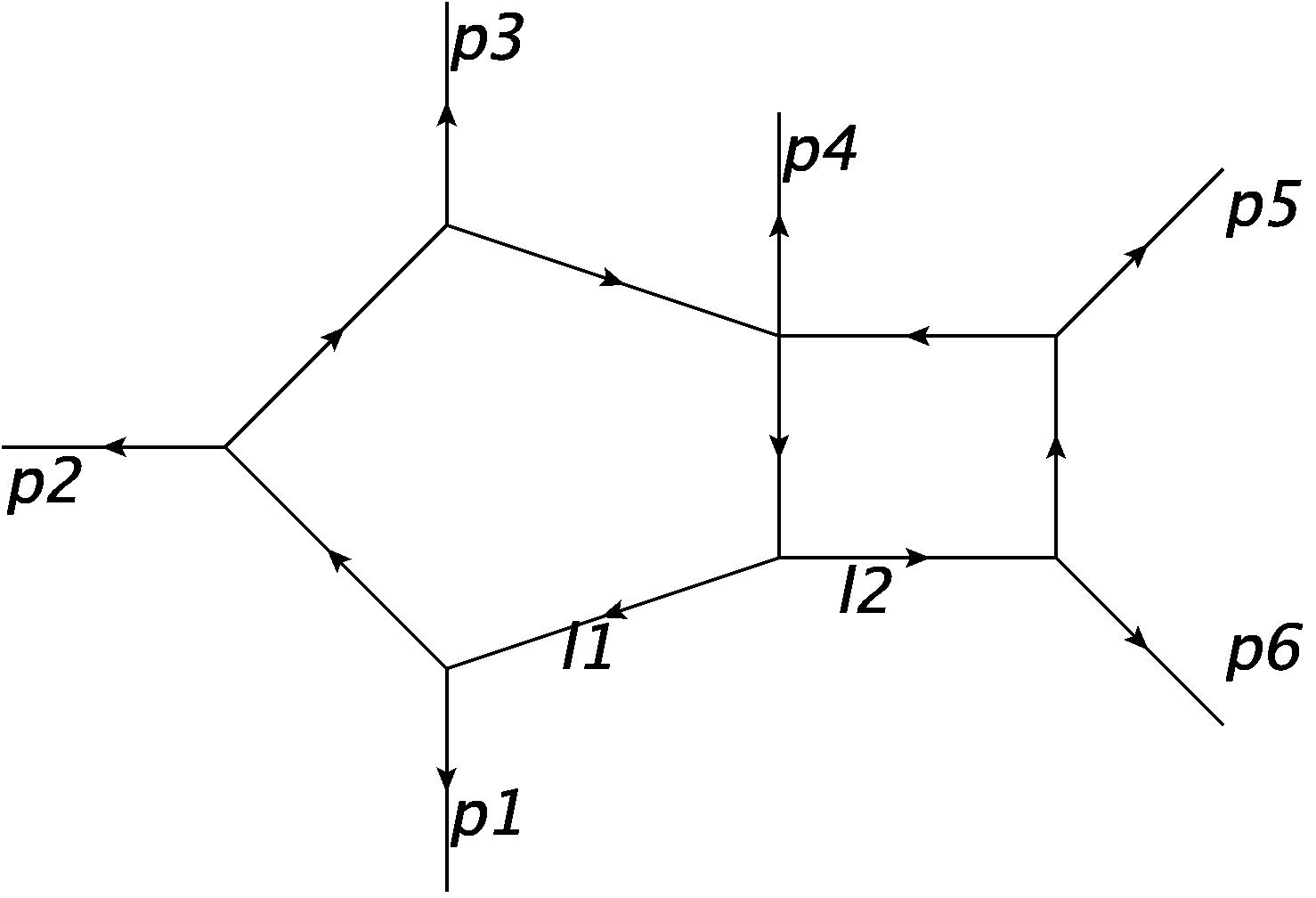}}
   \subfigure [Pentagon-triangle ] {\includegraphics[width=5 cm]{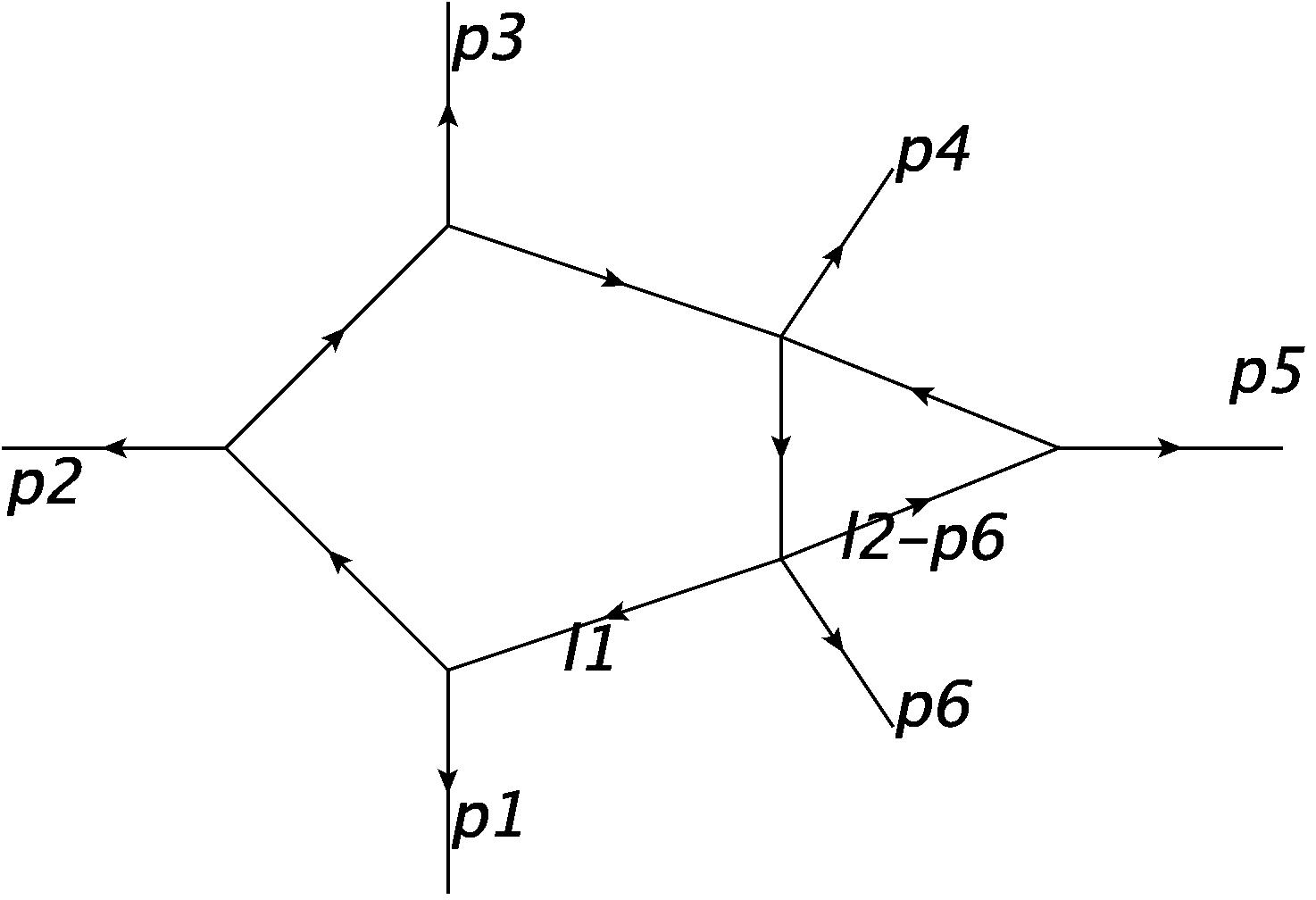}}
  \subfigure [Double-box] {\includegraphics[width=5 cm]{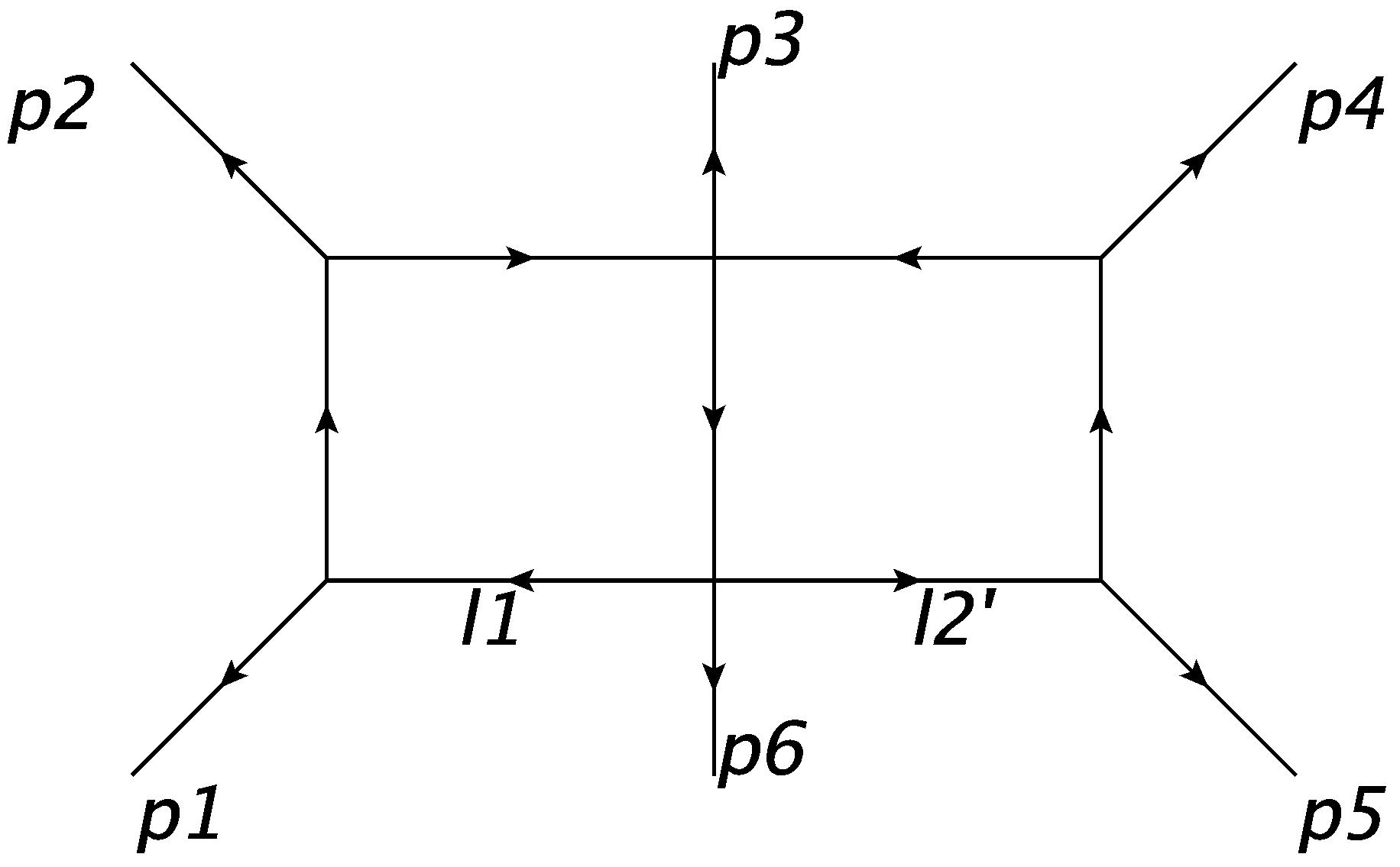}}
   \subfigure [Hexagon-box] {\includegraphics[width=4cm]{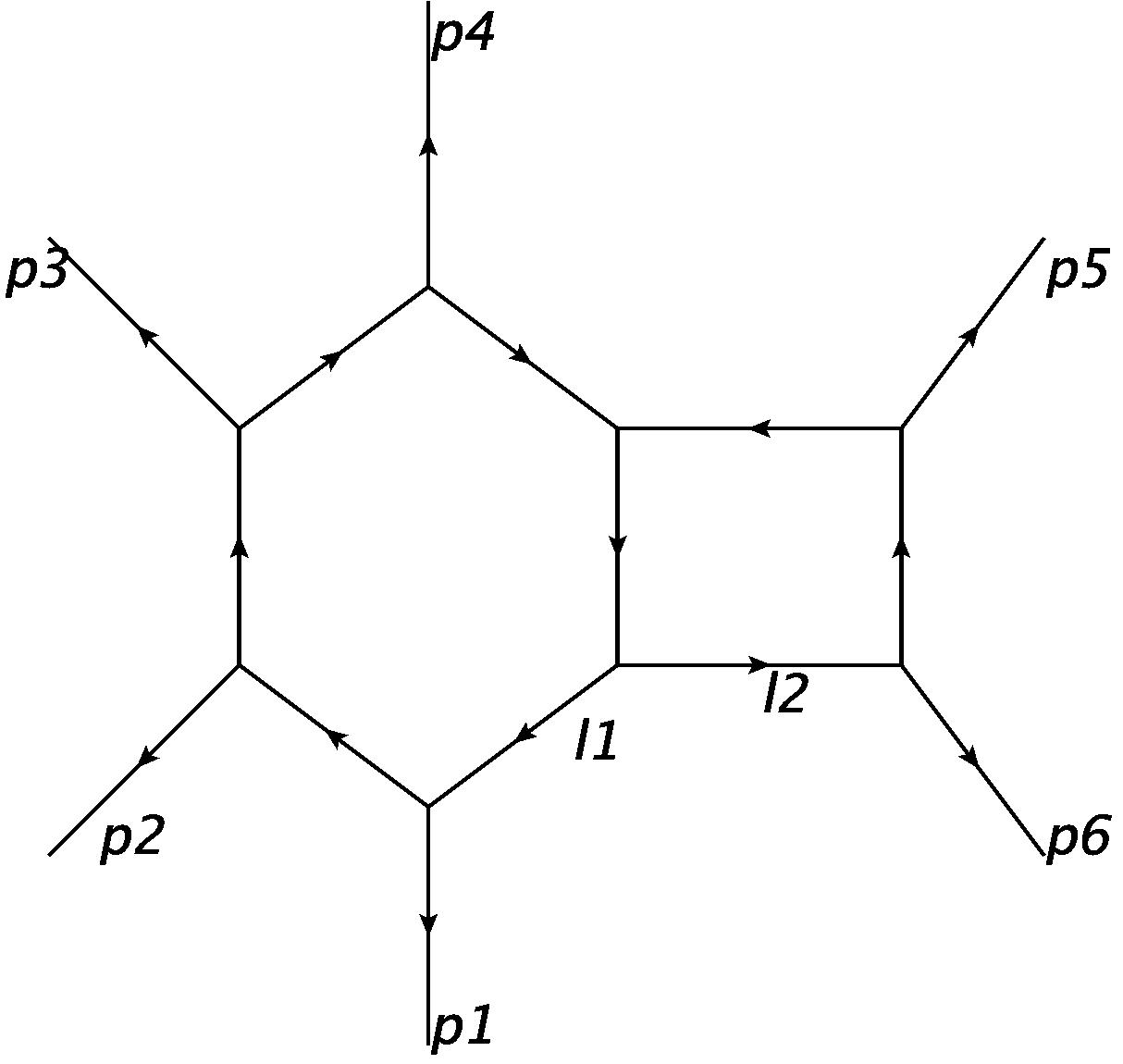}}
  \subfigure [Hexagon-bubble] {\includegraphics[width=3.2cm]{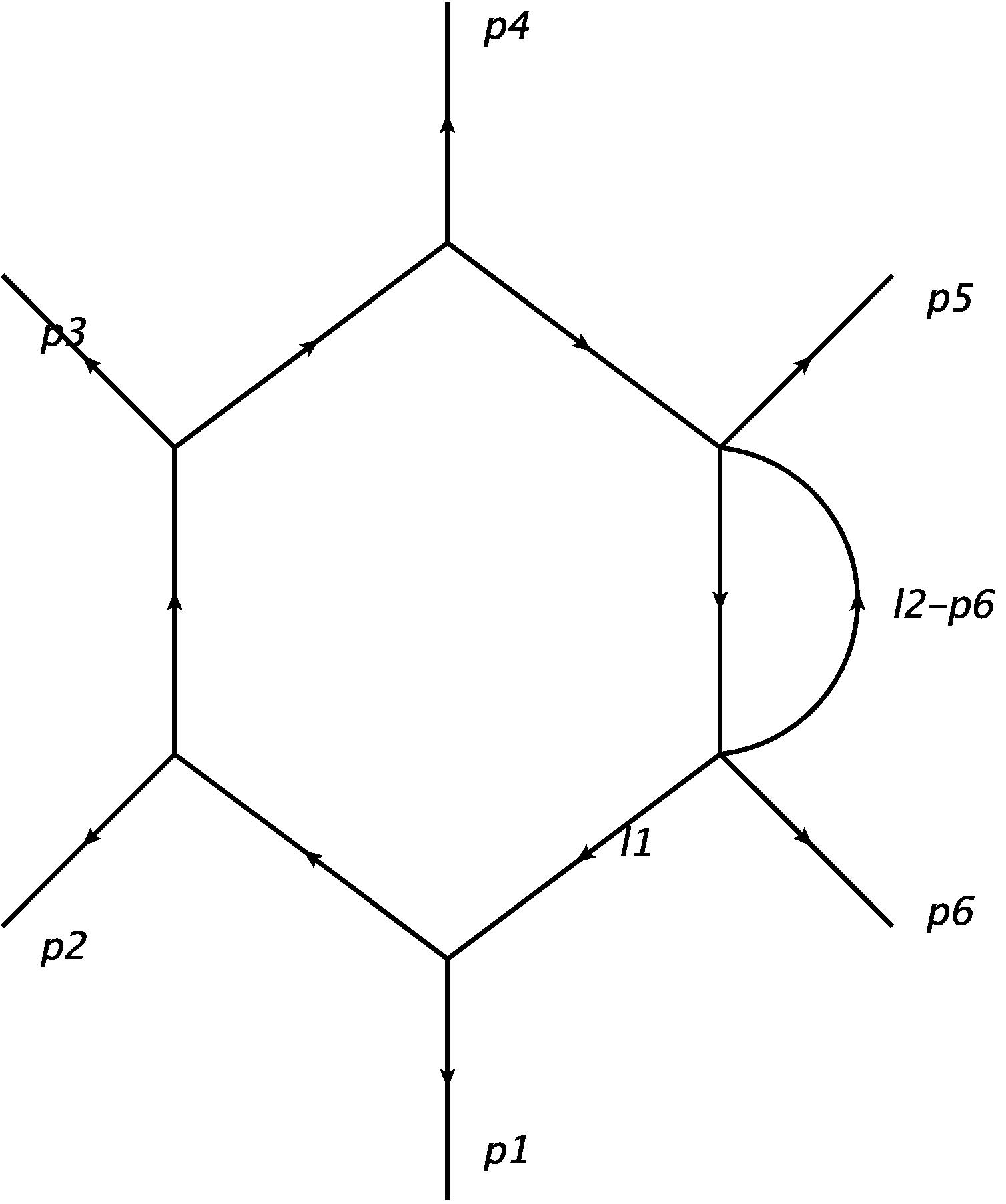}}
  \caption{Genuine two-loop six-point massless planar diagrams.}
\label{figure:Genuine}
\end{figure}

To determine the topologies of ``genuine'' two-loop
six-point diagrams, we generate the two-loop six-point diagrams
consisting only of three-point vertices, and then determine integral
families. We find there are three planar two-loop six-point integral
families, characterized by the top-sector diagrams corresponding to a double pentagon, 
a hexagon-box, and a heptagon-triangle integral.
We then analyze integral subsectors (obtained by pinching propagators), and
keep the ones that are genuine in the above sense.

Then, performing a numeric IBP analysis with {\sc Azurite},
we see that the master integrals from the double pentagon and hexagon-box families
(as well as those from two-loop five-point integrals with one off-shell leg), cover the master
integrals from the heptagon-triangle family. In other words, the heptagon-triangle family
does not provide new master integrals. Therefore, it is sufficient for us to focus on the
other two integral families.   
The final list of integral sectors we need to consider is shown in Fig.~\ref{figure:Genuine}.

\begin{table}[t]
\centering
\begin{tabular}{|c|c|c|c|c|c|c|}
\hline
Diagram & (a) & (b) &(c)
 & (d)&(e)&(f)\\\hline
Abbreviation &DP-a & DP-b &DP-c 
 & DP-d&HB-a&HB-b\\\hline
Number of propagators & 9 & 8 &7 & 7&9&7 \\\hline
Number of master integrals & 5&3 &1 & 7 & 1& 1 \\
\hline
\end{tabular}
\caption{Information for the genuine two-loop six-point massless
  planar diagrams. Here the master integer counting refers to the
  particular sector. }
\label{table:integral_family_info}
\end{table}

In order to perform IBP reduction, one needs to choose irreducible scalar products.
For the double pentagon (DP) and hexagon-box (HB) families, we define the following factors
\begin{gather}
D_1=l_1^2,D_2=\left(l_1-p_1\right){}^2,D_3=\left(l_1-p_1-p_2\right){}^2,D_4=\left(l_1-p_1-p_2-p_3\right){}^2,\nonumber
\\
D_5=\left(l_1+l_2\right){}^2,D_6=l_2^2,D_7=\left(l_2+p_1+p_2+p_3+p_4+p_5\right){}^2,D_8=\left(l_2+p_1+p_2+p_3+p_4\right){}^2,\nonumber
\\D_9=\left(l_2+p_1+p_2+p_3\right){}^2,D_{10}=\left(l_1+p_5+p_6\right){}^2,D_{11}=\left(l_2+p_1+p_2\right){}^2 \,.
\label{eq:Propagators}
\end{gather}
This allows us to treat all integrals using one common notation.
For example, the DP family corresponds to the sector $(1,1,1,1,1,1,1,1,1,0,0)$, while the HB family corresponds to $(1,1,1,1,1,1,1,1,0,1,0)$.

\section{Methodology}

\subsection{Canonical differential equations, cuts, and function alphabet}
\label{sec:canonicalDE}

The method of differential equations \cite{etde_5621240,Bern:1993kr,Gehrmann:1999as} is a state-of-the-art tool for computing Feynman integrals.
However, beyond the simplest cases, a naive application of the method can lead to prohibitive algebraic complexity. 
In references \cite{Henn:2013pwa,Henn:2014qga} it was argued that with a suitable choice
of master integrals, differential equations can be simplified to a canonical form.
The latter can be solved easily order by order in a Laurent series expansion in $\epsilon$, the dimensional regularization parameter. 
The canonical form of the differential equation one can hope to reach is
\begin{gather}
    d\vec{f}=\epsilon (d \tilde A) \vec{f}\,.
\label{DEcanonical}
\end{gather}
In the case of multiple polylogarithms, $\tilde A$ takes the form  
\begin{gather}
    \tilde A=\sum a_k \log W_k \,,
\label{dlog}
\end{gather}
where each $a_k$ is a constant matrix, and the $W_k$'s are functions of Lorentz invariants.
The individual $W_{k}$ are called letters, and the set $\{ W_{k} \}$ is called alphabet.
The significance of these letters is that they specify the function space (and, as a proxy, the symbol \cite{Goncharov:2010jf} of the functions) necessary to express the solutions to the differential equations.

Let us comment on the solution to the differential equations. 
If we write the solution as a Laurent series,
\begin{align}
   \epsilon^{2L} \vec{f}=\vec{f_0}+\epsilon \vec{f_1}+ \epsilon^2 \vec{f_2}+\dots\,,
\end{align}
then, thanks to the proportionality of the R.H.S. of (\ref{DEcanonical}) to $\epsilon$, the differential equations decouple order by order in $\eps$,
\begin{align}
\label{UT-expansion}
    d \vec{f_1}=(d \tilde A) \vec{f_0}\,, \qquad
     d \vec{f_2}= (d \tilde A) \vec{f_1} \,, \dots \,.
\end{align}
Therefore, $f_{k}$ is given by $k$-fold iterated integrals, with the integration kernels being elements of the matrix $d\tilde A$.
We remark that in most cases, the boundary information can be obtained from physical consistency conditions, see e.g. \cite{Chicherin:2021dyp}.

To summarize, if we choose our basis integrals $\vec{f}$ in a judicious way, the
resulting differential equations are simplified a lot, ideally to the form (\ref{DEcanonical}),
which is easily solved.

How should the basis integrals be chosen? The key property they should have
can be described with the concept of the degree of `transcendentality', or transcendental weight,
$\mathcal{T} (f)$ of a function. 
We define $\mathcal{T}(\mathrm{Li}_k(x))=k$ since $\mathrm{Li}_k(x)$ is a $k$-fold iterated integral. Moreover, we require $\mathcal{T}(f_1 f_2)=\mathcal{T}(f_1)+\mathcal{T}(f_2)$.  Therefore we have
\begin{gather}
    \mathcal{T}(\log x)=1, \quad\mathcal T(\pi)=1,\quad \mathcal{T}(\zeta_n)=\mathcal{T}(\mathrm{Li}_n(1))=n,\\\nonumber
    \mathcal{T}(\text{algebraic  factors})=0,\quad \mathcal{T}(\zeta_2)=\mathcal{T}(\frac{\pi^2}{6})=2.
\end{gather}
An integral $J$ has the uniform transcendental (UT) weights if in the
expansion $\epsilon^{2L} J= \sum_{n=0}^\infty \epsilon^n J_n$,
\begin{equation}
  \label{eq:1}
  \mathcal T(J_n)=n
\end{equation}

If a function $f$ also satisfies 
\begin{gather}
    \mathcal{T}\left(f'\right)=\mathcal{T}(f)-1\,,
\end{gather}
where the prime stands for any partial derivative, then the function $f$ is called a pure function. 
This last criterion assures that one deals with a $\mathbb{Q}$-linear combinations of uniform weight functions, as opposed to functions with kinematic-dependent prefactors. 

The usefulness of these concepts is that pure uniform weight functions satisfy simple differential equations.
In particular, if all basis elements $I$ are pure uniform weight functions, then the corresponding differential equation is canonical \cite{Henn:2013pwa}.
Therefore being able to identity pure uniform weight integrals is paramount. We outline our approach to this problem in section \ref{Baikov}.

Let us discuss a structural property of the differential equations (\ref{DEcanonical}) that will be important in the following. 
The propagator structure of Feynman integrals implies a hierarchy according to which the integrals $f$ can be arranged.
Specifically, if an integral corresponds to a subsector of another integral (i.e. it has only a subset of propagator factors in its definition), this implies a zero in an off-diagonal element of the differential equation matrix $\tilde{A}$.
It is therefore useful to define integral sectors, corresponding to particular propagator structures, such as the ones shown in Fig.~\ref{figure:Genuine}, and their subsectors. When this is done, $\tilde{A}$ takes a block form. As mentioned earlier, all blocks corresponding to five-point integrals with one off-shell leg have already been computed in previous work. 
What remains to be done is to compute the blocks corresponding to the sectors in Fig.~\ref{figure:Genuine} (these are blocks arranged along the diagonal), as well as off-diagonal blocks that connect their derivative to the previously known integrals.

Experience shows that obtaining the canonical form (\ref{DEcanonical}) of the differential equations is the most complicated step. The reason is that modifications to the basis integrals $\vec{f}$ needed to take care of subintegrals have a special matrix form and can, at least in principle, be obtained algorithmically \cite{Caron-Huot:2014lda,Gehrmann:2014bfa}.
In this paper we focus on the diagonal blocks for all genuine six-particle sectors shown in Fig.~\ref{figure:Genuine}.
We present an integral basis that puts the differential equation corresponding to these sectors into canonical form,
and we identify novel alphabet letters $W_k$ that appear in those differential equations.

\subsection{Integrand analysis to find uniform weight integrals} 
\label{Baikov}

One crucial problem of using differential equation to solve Feynman
integrals, is to find a UT basis. 
There are different approaches to this, which can be classified into two types. 
Firstly, a number of approaches have been developed that aim at first computing the differential equations matrix, in any basis, 
and then simplifying it using appropriate transformations. A guiding principle in finding the necessary transformations is the expected
singularity structure \cite{Henn:2014qga,Lee:2014ioa}. There are a number of computer codes dedicated to this \cite{Meyer:2016slj, Meyer:2017joq,Prausa:2017ltv, Gituliar:2017vzm,Lee:2020zfb,Dlapa:2020cwj}.
Secondly, another approach uses insights at the loop integrand level to identify UT integrals. In particular, integrands with only logarithmic singularities, and kinematic-independent maximal residues (leading singularities) that are expected \cite{ArkaniHamed:2010gh,Henn:2013pwa} to be UT functions. These ideas have been automated in \cite{Wasser:2018qvj,Henn:2020lye}.

The second approach has proven extremely useful, especially for Feynman integrals with many kinematic variables or several singular points. It has the advantage that, if successful, the differential equation matrix that needs to be computed is drastically simplified, thereby avoiding an enormous amount of algebraic complexity. Even if one only 'almost' obtains a canonical form, it is then usually much easier to find the remaining transformations. 

Let us therefore review the integrand analysis.
It was noticed, first in  $\mathcal N=4$ super-Yang-Mills
amplitudes~\cite{ArkaniHamed:2010gh},
that integrals with constant four-dimensional leading singularities 
(and whose integrands have only single poles) give rise to UT
integrals. It was later seen that this principle also applies to
integrals in other theories, and as such plays an important role
in the differential equations method \cite{Henn:2013pwa}.

Consider an $L$-loop integral, suppose the dimension of loop momenta
is $4$.  The leading singularities of an integral~\cite{Cachazo:2008vp} are obtained by evaluating 
the integrand on certain integration contours around (multiple) poles of the
integrand. Schematically, the contour integral
has the form
\begin{gather}\label{example4dleadingsingularity}
    \frac{1}{(2 \pi i )^{4L}}\oint d^4
    l_1\ldots d^4 l_L \frac{N}{D_1 \ldots D_k}\, .
\end{gather}
For example, if the number of propagators $k$, equals the fold of integration
$4L$, then one type of leading singularity is obtained by taking $\oint_{D_1 =0, \ldots, D_k=0}$,
the so-called maximal cut. Then the leading singularity can be simply calculated by
Cauchy's formula. In general, there may be other leading singularities due to
additional poles that originate from Jacobian factors.
In this case, or when $k<4L$, the leading singularity may be computed
by recursive residue computations (which include the so-called
composite leading singularities). For complicated multivariate residue
computations, algebraic geometry techniques like the
transformation law and Gr\"obner basis can be very
useful~\cite{Sogaard:2013fpa,Larsen:2017kzf}.
The upshot is that for a given propagator structure as in (\ref{example4dleadingsingularity}), the question which integrals (i.e., which loop-dependent numerators $N$) 
lead to integrands with unit leading singularities, can be answered algorithmically \cite{Wasser:2018qvj,Henn:2020lye}. In a very recent paper~\cite{Bourjaily:2021hcp}, a four-dimensional basis of dlog integrands for planar six-particle two-loop integrands was constructed.

As mentioned in the introduction, in some cases, especially those with many external scales \cite{Chicherin:2018old},
the four-dimensional integrand analysis may not be enough.
In such cases it is natural to treat the integrand in $D$ dimensions.

In particular, it has proven useful to analyze integrands in the Baikov representation.
In this way, one treats some of the $D$-dependence exactly, while still using the same ideas about the singularity structure of the integrand (but now in a different representation) \cite{Chicherin:2018old, Dlapa:2021qsl}.
(see also \cite{Chen:2020uyk}, where techniques of intersection theory are used to develop Baikov dlog forms for UT integrals.)
The $D$-dimensional Baikov analysis reveals some hidden information
not seen from the 4D leading singularities. An integral may have all
$4D$ residues to be constants, but some non-constant $D$-dimensional residues \cite{Chicherin:2018old}. 
In this paper, we use the four-dimensional analysis as a starting point, and then refine and complement it by the Baikov D-dimensional leading singularity
method.  Let us briefly review the Baikov representation method.

Here we briefly introduce the Baikov representation~\cite{Baikov:1996cd},
and give an example of the leading singularity analysis in Baikov presentation.

In the scheme of dimensional regularization, 
\begin{gather}
\label{Feynman integral-1}
    I(\alpha_1,...,\alpha_m;D)\equiv\int\prod_{j=1}^L\frac{d^D l_j}{(i\pi)^{D/2}}\frac{1}{D_1^{\alpha_1}...D_m^{\alpha_m}},
\end{gather}
here external momenta are $p_1,...,p_E,p_{E+1}$ and the loop momenta
are $l_1,...,l_L$. The propagators are labeled so that
\begin{align}
        &\alpha_i\ge1,  \quad    i=1,...,k\\
        &\alpha_i\le0,   \quad  i=k+1,...,m\,.
\end{align}

Next, define a set of all independent external and loop momenta:
\begin{equation}
    {v_1,...,v_{E+L}} = {p_1,...,p_E,l_1,...,l_L}\, ,
\label{momenta}
\end{equation}
and form their Gram matrix $S$,
\begin{equation}
    S_{ij}=v_i\cdot v_j .
\end{equation}
And we can define another Gram matrix $G$, with
\begin{equation}
   G_{ij}=v_i\cdot v_j, \quad 1\le i,j\le E \,. \label{G-matrix}
\end{equation}
From the product of momenta in \eqref{momenta}, we can get $m$ scalar products
\begin{equation}
  m=LE+\frac{L(L+1)}{2}  \, .
\end{equation}

The Baikov representation uses the inverse propagators and irreducible scalar products as variables,
\begin{gather}
    z_{\alpha}\equiv D_{\alpha},\quad 1\le \alpha \le m\,.
\end{gather}
The Jacobian associated with the change of variables
from $(l_1^\mu,..,l_L^\mu)$ to $(z_1, . . . , z_m)$ is an appropriate
power of the determinant of $S$,
\begin{equation}
    S\equiv \det S_{ij}\, .
\label{S-matrix}
\end{equation}

The Baikov representation of the integral \eqref{Feynman integral-1} is given by the formula
\begin{gather}
       I(\alpha_1,...,\alpha_m;D)=C_{L}^E(D)S_{E}^\frac{E-D+1}{2}\int\frac{dz_1...dz_m}{z_1^{\alpha_1}...z_m^{\alpha_m}}
       S^\frac{D-L-E-1}{2}\, ,
\label{Baikov_integrand}
\end{gather}
where the first prefactor $C_{L}^E(D)$ is just a function of dimension $D$, which will not influence our discussion. Here $S_{E}=\det G$ in \eqref{G-matrix} and $S$ is defined in \eqref{S-matrix}.

The Baikov leading singularities are the multivariate residues of the
expression \eqref{Baikov_integrand}. The advantage of this kind of
leading singularity analysis is that the formula benefits from the
explicit $D$-dependence and the simple structure of the denominators.

In practice, we often use the loop-by-loop Baikov representation~\cite{Frellesvig:2017aai,Harley:2017qut}. This
approach
can often reduce the number of variables related to ISPs (irreducible
scalar products). We  use an example to
explain how to do the  loop-by-loop Baikov leading singularity analysis.

\subsubsection*{Example: Hexagon-box diagram in Baikov parametrization}
\label{example:hb-a}

The hexagon-box integral sector HB-a is shown in Fig.~\ref{figure:Genuine}(e). There is one master integral in this sector. The propagators are written in Eq\eqref{eq:Propagators}.

First, treat the right part as a four point box diagram with
external momenta $p_5$, $p_6$, $l_1$ and loop momenta $l_2$ with four
propagators . We can write a $D$-dimensional Baikov representation of the right part, with the following $S$ and $S_E$,
\begin{gather}
S_R=G\left(
\begin{array}{cccc}
 l_2 & p_5 &  p_6 & l_1 \\
 l_2 & p_5 &  p_6 & l_1 \\
\end{array}
\right), \quad
   S_{E,R}= G\left(
\begin{array}{ccc}
 p_5 & p_6 & l_1 \\
 p_5 & p_6 & l_1 \\
\end{array}
\right).
\end{gather}

Then consider the left loop. The remaining left part is a pentagon diagram with external momenta $p_1, p_2, p_3, p_4$, loop momentum $l_1$ with the remaining five propagators. Also, we have
\begin{gather}
    S_{E,L}=G\left(
\begin{array}{cccc}
 p_1 & p_2 & p_3 & p_4 \\
 p_1 & p_2 & p_3 & p_4\\
\end{array}
\right),\quad
S_L=G\left(
\begin{array}{ccccc}
 l_1 & p_1 & p_2 & p_3 & p_4 \\
 l_1 & p_1 & p_2 & p_3 & p_4 \\
\end{array}
\right).
\end{gather}
 
The loop-by-loop Baikov
representation for HB-a is
\begin{gather}
  \label{eq:27}
  \int dz_1\ldots z_{9} \ G\left(
\begin{array}{cccc}
 {p_1} & p_2 &p_3& {p_4} \\
 {p_1} & p_2 &p_3& {p_4} \\
\end{array}
\right)^\frac{5-D}{2} G\left(
\begin{array}{ccccc}
l_1 & {p_1} & p_2 &p_3& {p_4} \\
l_1 & {p_1} & p_2 &p_3& {p_4} \\
\end{array}
\right)^\frac{D-6}{2} \nonumber \\
\times 
G\left(
\begin{array}{ccc}
 p_5 &p_6& {l_1} \\
 p_5 &p_6& {l_1} \\
\end{array}
\right)^\frac{4-D}{2} 
G\left(
\begin{array}{cccc}
l_2 & p_5 &p_6& {l_1} \\
l_2 & p_5 &p_6& {l_1} \\
\end{array}
\right)^\frac{D-5}{2} \times \frac{N}{D_1 \ldots D_9}\,,
\end{gather}
where the Baikov variables $z_1,\ldots z_9$ are
\begin{gather}
  \label{eq:30}
  l_2^2, \quad l_2 \cdot p_5,\quad l_2 \cdot p_6,\quad l_2 \cdot
  l_1\nonumber \\
l_1^2,\quad l_1 \cdot p_1,\quad l_1 \cdot p_2,\quad l_1 \cdot p_3,
\quad l_1 \cdot p_4\,.
\end{gather}

Under the maximal cut $D_1=\dots=D_9=0$, and set $D=4$, the Baikov residue is
\begin{gather}
    \text{Res}(I)= s_{56}^{-1} (l_{1}\cdot p_6)^{-1}(S_{E,L})^{1/2} (S_{L})^{-1} N_{\text{maxcut}}\, .
\end{gather}

From this expression, we can construct a following candidate numerator
\begin{align}
  N_1^\text{HB-a}=\frac{G\left (
  \begin{array}{ccccc}
    l_1& p_1 & p_2 & p_3 & p_4 \\
 l_1& p_1 & p_2 & p_3 & p_4 
  \end{array}
\right)}{G\left(\begin{array}{cccc}
    p_1 & p_2 & p_3 & p_4 \\
 p_1 & p_2 & p_3 & p_4 
  \end{array}
\right)^{1/2}} s_{56} (l_1+p_6)^2\,.
\label{HB-a:UT}
\end{align}
The fraction between two Gram determinants
cancels the residue from the left loop in the Baikov representation,
while the factor $s_{56}(l_1+p_6)^2$ cancels the residue from the
right box loop. The numerator thus has constant
leading singularity in Baikov representation.

\subsection{Finite field IBP reduction and reconstruction of differential equations}

We use both FIRE6~\cite{Smirnov:2019qkx} and the
finite-field computational framework \textsc{FiniteFlow}~\cite{Peraro:2016wsq,Peraro:2019svx} for the IBP reduction in this
paper. As mentioned in section \ref{sec:kinematics}, because of the four-dimensional
spacetime of the external states, only four of the six external momenta can be linearly
independent. Using the momentum-twistor parametrization defined in~\eqref{eq:t3} we represent the six external momenta in the propagators as linear combinations of four external momenta, with coefficients that are rational functions of momentum twistor variables.

The analytic IBP reduction using eight variables
is challenging. When a numeric IBP reduction is needed to check if a
differential equation is canonical, one can set the eight
momentum twistor variables to be random integers to get
the numeric reduction table.  As mentioned at the end of section~\ref{sec:canonicalDE}, in this paper we focus on the diagonal blocks of the differential equations that correspond to six-particles sectors.  To obtain these diagonal blocks of the differential equations, for each sector, we perform the reductions on its maximal cut, i.e.\ setting integrals of lower subsectors
 to zero.  This makes a numerical IBP reduction feasible with either FIRE6 or \textsc{FiniteFlow}.

However, a numeric IBP reduction is not sufficient to derive the analytic
differential equations, which we need to obtain analytic expressions of symbol
letters $W_k$ after integrating them into the form of equation~\eqref{dlog}. For this purpose, we resort to
 finite fields and functional reconstruction methods, implemented in the program {\sc
  FiniteFlow} \cite{Peraro:2016wsq,Peraro:2019svx}.  We follow the strategy described in ref.~\cite{Peraro:2019svx} to directly reconstruct the differential equation matrices on the cuts (without the need of reconstructing more complicated analytic IBP tables).
  
  The  derivatives of the Feynman integrals
in the momentum twistor variables can be obtained as proper linear combinations of
derivatives in the external momenta.  These derivatives are thus reduced to master integrals again, to obtain the differential equation matrices.  The system of IBP identities (which also includes symmetries and Lorentz invariance identities) is generated with the help of the package {\sc LiteRed}~\cite{Lee:2013mka} and thus reduced numerically (on the cut) to master integrals using \textsc{FiniteFlow}.  The reduction is performed over finite fields of integers modulo a machine size prime.  This yields a numerical evaluation of the differential equation matrices.  Full analytic expressions for the differential equations are thus reconstructed using the methods in~\cite{Peraro:2016wsq,Peraro:2019svx}, from repeated numerical evaluations with different numerical inputs for the momentum twistor variables and for the prime that defines the finite field.  Using this strategy the diagonal blocks of the differential equation matrices can be found with very modest computing resources.

A  subtlety comes from the fact that some UT integrals have a square root prefactor in their
definition, whose analytic expression comes from the
Baikov analysis.  As shown in ref.~\cite{Peraro:2019svx}, one can still perform the reconstruction of the full differential equations without performing any non-rational operation. If $R$ is the square root of a rational function, then $\frac{R^{'}}{R}$ is rational. This implies that, via IBP identities, the correct square-root factors can thus be easily restored at the end (for more details see~\cite{Peraro:2019svx}).

\section{Results for uniform weight integrals}

In section \ref{sec:integral-families} we discussed the genuine planar two-loop six-particle integral sectors that we analyzed.
They are shown in Fig.~\ref{figure:Genuine}. Table~\ref{table:integral_family_info} contains the notation and the number of master integrals.
In this section we describe the outcome of our leading singularity and Baikov analysis.
For each integral sector, we provide a basis of master integrals that leads to canonical differential equations (on the cut).
Moreover, we identify the alphabet letters appearing in the cut differential equations, and in each case remark which new letters are encountered.
For convenience of readers, all these results are available as computer-readable ancillary files.

\subsection{Double pentagon}
\label{subsection:double pentagon}
Here we discuss the double pentagon sector shown in Figure~\ref{figure:Genuine}(a).
Recall that there are five master integrals in this sector.

We begin by a four-dimensional leading singularity analysis of the octa-cut.
We find the following that the following numerators lead to integrands with constant leading singularities.
\begin{align}
  N_1^\text{DP-a}&= (w_2+v_2)^2 s_{13} s_{46} (l_1- w_1)^2 (l_2- v_1)^2\, \label{dpa_N1} \,, \\
 N_2^\text{DP-a}&=  (w_2+v_1)^2 s_{13} s_{46} (l_1- w_1)^2 (l_2- v_2)^2\, ,\label{dpa_N2} \\
N_3^\text{DP-a}&= (w_1+v_2)^2 s_{13} s_{46} (l_1- w_2)^2 (l_2- v_1)^2 \, ,\label{dpa_N3} \\
N_4^\text{DP-a}&= (w_1+v_1)^2 s_{13} s_{46} (l_1- w_2)^2 (l_2- v_2)^2 \, .\label{dpa_N4} 
\end{align}

Here, with the notation of spinor helicity notations,

\begin{align}
  \label{eq:a3}
  w_1&=p_1+\frac{[23]}{[13]}\lambda_2 \tilde \lambda_1,\quad w_2= w_1^* \,,\\
 v_1&=p_6+\frac{[54]}{[64]}\lambda_5 \tilde \lambda_6,\quad v_2= v_1^* \,.
\end{align}

These terms are reminiscent of the chiral numerators first introduced
in~\cite{ArkaniHamed:2010gh}. All these four integrals are finite
integrals. We remark that the integral with the
numerator $N_1^\text{DP-a}$ in \eqref{dpa_N1}, in the limit $\epsilon
=0$, is the double pentagon
integral with two wiggly lines defined in
ref.~\cite{ArkaniHamed:2010gh}. Similarly, the integral with
$N_4^\text{DP-a}$ in \eqref{dpa_N4}, in the limit $\epsilon
=0$, is the double pentagon
integral with two dashed lines defined in
ref.~\cite{ArkaniHamed:2010gh}. The $\epsilon=0$ limit of the two rest
integrals in \eqref{dpa_N2} and in \eqref{dpa_N3} are the
 integrals with mixed numerators. 

The four-dimensional integration of these numerators are analytically
calculated in ref.~\cite{Drummond:2010cz, Dixon:2011nj}. The 4D
integrals of the numerators $N_1^\text{DP-a}$ and $N_4^\text{DP-a}$
equal each other. However, the $O(\epsilon)$ parts of these integrals are
different. In the dimensional regularization scheme, the four integrals with
the numerators $N_i^\text{DP-a}$, $i=1,2,3,4$, are linearly independent by the
IBP analysis.

It is not easy to construct integrals with other forms of chiral numerators
such that  they (1) have constant 4D leading singularity, (2) and 
are independent of the integrals defined in \eqref{dpa_N1} to \eqref{dpa_N4} by IBP relations.  Note that there are $5$ master
integrals on this sector, so we would like to find more UT candidates
by considering integrals with $5\times 5$ Gram determinant numerators and
integrals in 6D, like the strategy in
the ref.~\cite{Chicherin:2018old, Abreu:2018aqd}.

We remark that although integrals in \eqref{dpa_N1} to \eqref{dpa_N4} have  constant four-dimensional leading
singularities, in dimensional regularization they are not guaranteed
to be UT integrals. However, as suggested in the
ref.~\cite{Chicherin:2018old}, such kind of integrals can be upgraded to
UT integrals with the addition of certain $5\times 5$ Gram determinant
integrals.

We determine the UT candidates on this sector from
$D$-dimensional Baikov leading singularity computations. 
The resulting five integrals are (a more detailed explanation follows below)
\begin{align}
  \label{eqDoublePentagonBaikov1}
  I_1^\text{DP-a} & =\int \frac{d^{4-2\eps} l_1}{i \pi^{2-\eps}} \frac{d^{4-2\eps} l_2}{i
  \pi^{2-\eps}} \frac{N_1^\text{DP-a}-N_4^\text{DP-a}}{D_1 \ldots D_9} \, , \\
  I_2^\text{DP-a} & =\int \frac{d^{4-2\eps} l_1}{i \pi^{2-\eps}} \frac{d^{4-2\eps} l_2}{i
  \pi^{2-\eps}} \frac{N_2^\text{DP-a}-N_3^\text{DP-a}}{D_1 \ldots D_9}  \, ,\\
 I_3^\text{DP-a} & =F_3\int \frac{d^{4-2\eps} l_1}{i\pi^{2-\eps}} \frac{d^{4-2\eps} l_2}{i
 \pi^{2-\eps}} \frac{\mu_{12}
}{D_1 \ldots D_9} \, , \label{dpa_UT_3}
\\
I_4^\text{DP-a} & =F_4 \epsilon^2 \int \frac{d^{6-2\eps} l_1}{i\pi^{3-\eps}}
                  \frac{d^{6-2\eps} l_2}{i\pi^{3-\eps}} \frac{1}{D_1
                  \ldots D_9}  \, , \label{dpa_UT_4}\\
I_5^\text{DP-a} & =\int \frac{d^{4-2\eps} l_1}{i \pi^{2-\eps}} \frac{d^{4-2\eps} l_2}{i
  \pi^{2-\eps}} \frac{N_1^\text{DP-a}+N_4^\text{DP-a}+F_5\mu_{12}}{D_1 \ldots D_9} \, , \label{eqDoublePentagonBaikov5}
\end{align}
where 
\begin{equation}
  \label{eqBaikovDoublePentagonGram}
 \mu_{12}= \frac{G\left(
                   \begin{array}{ccccc}
l_1& p_1& p_2& p_3& p_6\\
l_2& p_1& p_2& p_3& p_6
                   \end{array}\right)}{G(1,2,3,6)}\,.
\end{equation}

The following comments are in order:
\begin{itemize}
\item The first two UT integrals $I_1^\text{DP-a}$ and
  $I_2^\text{DP-a}$ simply correspond to the two parity odd
  combinations of the four integrals of
  Eqs. (\ref{eqDoublePentagonBaikov1}) -
  Eqs. (\ref{eqDoublePentagonBaikov5}). As pointed out in
  \cite{Drummond:2010cz},  $I_1^\text{DP-a}$  has vanishing $O(\epsilon^0)$
  order. However, the nonvanishing $O(\epsilon)$ order makes it an
  indepedent integral.

\item The third UT $I_3^\text{DP-a}$ has the numerator given in Eq. (\ref{eqBaikovDoublePentagonGram}). The latter is a  $5\times 5$ determinant that involves the loop momenta. As a consequence, it vanishes for four-dimensional loop momenta, but has a non-trivial Baikov leading singularity, which is given by $F_3$. 
$F_3$ is a rational function of $s_{ij}$ divided by $\epsilon_{1236}$, but its expression is rather long.
We were able to find a rather compact representation for it in spinor notation,
\begin{equation}
  \label{dpa_F3a}
  F_3={(\la 12 \ra [23] \la 34 \ra [45] \la 56 \ra [61]-\la 23 \ra [34] \la
45 \ra [56] \la 61 \ra [12]) s_{456}}/{8 } \,.
\end{equation}
In addition, with the help of finite field methods, we found the
following relation to the alphabet letters given in the subsection \ref{subsection:even_letters}, namely $F_3 = {W_{46} W_{145}}/{8}$.

\item The fourth UT is the six-dimensional six-point double pentagon. Here the
  internal momenta are $6-2\epsilon$ dimensional, while the external
  momenta are still in four dimensions. 
  
  Let us sketch how to find the
  coefficient for this UT integral.
By a loop-by-loop Baikov representation, the integral has the
representation,
\begin{gather}
\int d^9z \ G\left(
                   \begin{array}{ccccc}
l_2 &l_1& p_4& p_5& p_6\\
l_2 &l_1& p_4& p_5& p_6
                   \end{array}\right)^{\frac{D-6}{2}} 
G\left(
                   \begin{array}{cccc}
l_1& p_4& p_5& p_6\\
l_1& p_4& p_5& p_6
                   \end{array}\right)^{\frac{5-D}{2}} \nonumber \\
\times G\left(\begin{array}{ccccc}
l_1 &p_1& p_2& p_3& p_6\\
l_1 &p_1& p_2& p_3& p_6
                   \end{array}\right)^{\frac{D-6}{2}} 
G\left(
                   \begin{array}{cccc}
p_1& p_2& p_3& p_6\\
p_1& p_2& p_3& p_6 \end{array}\right)^{\frac{5-D}{2}} \,,
\end{gather}
where we dropped overall factors in $\epsilon$ only. The Baikov
variables are chosen to be
\begin{gather}
  l_2^2, \quad l_2\cdot p_4,\quad  l_2\cdot p_5,\quad l_2\cdot
  p_6,\quad l_2 \cdot l_1\nonumber \\
l_1^2, \quad l_1\cdot p_1,\quad l_1 \cdot p_2, \quad l_1 \cdot p_3, \quad l_1 \cdot p_6.
\end{gather}
Because of the loop-by-loop Baikov representation, we only have $10$
Baikov variables.

Taking the limit
$D\to 6$, two Gram determinants can be ignored for the leading
singularity. Consider the maximal cut with $D_1=\ldots =D_9=0$, then
only one fold of integral remains there,
\begin{gather}
  \text{6D leading singularity} =G\left(
\begin{array}{cccc}
p_1& p_2& p_3& p_6\\
p_1& p_2& p_3& p_6 
\end{array}\right)^{\frac{5-D}{2}} 
\int d(l_1\cdot p_6) \ G\left(
                   \begin{array}{cccc}
l_1& p_4& p_5& p_6\\
l_1& p_4& p_5& p_6
                   \end{array}\right)^{-\frac{1}{2}}\, .
\label{6dLS_max_cut}
\end{gather}
Here we impose the cut condition
\begin{gather}
  l_1^2=0,\quad l_1 p_1=0,\quad l_1 p_2=\frac{s_{12}}{2},\quad l_1 p_3=\frac{s_{12}+s_{13}+s_{23}}{2}.
\end{gather}
To evaluate \eqref{6dLS_max_cut}, it is convenient to set $p_1$, $p_2$,
$p_3$ and $p_6$ as a 4D basis, and expand the vectors $p_4$ and $p_5$
over this basis,

\begin{align}
p_4=c_{41}p_1 +c_{42}p_2+c_{43}p_3+c_{46}p_6 \equiv p_4^\sharp
  +c_{46}p_6\, ,\\
p_5=c_{51}p_1 +c_{52}p_2+c_{53}p_3+c_{56}p_6 \equiv p_5^\sharp
  +c_{56}p_6\, ,
\end{align}
where the notation ${}^\sharp$ means the component of a vector in the
linear subspace
$\text{span}\{p_1,p_2,p_3\}$. Explicitly,
\begin{equation}
p_4^\sharp =p_4 -\frac{\epsilon(1,2,3,4)}{\epsilon(1,2,3,6)} p_6 \,.
\end{equation}

By momentum conservation
$p_5^\sharp=-p_{123}-p_4^\sharp$. Then, using basic linear algebra,
the equation \eqref{6dLS_max_cut} can be simplified
as 
\begin{gather}
  \text{6D leading singularity} =G\left(
\begin{array}{cccc}
p_1& p_2& p_3& p_6\\
p_1& p_2& p_3& p_6 
\end{array}\right)^{\frac{5-D}{2}} 
\int d(l_1\cdot p_6) \ G\left(
                   \begin{array}{cccc}
l_1& p_4^\sharp& p_{123} & p_6\\
l_1& p_4^\sharp& p_{123}& p_6
                   \end{array}\right)^{-\frac{1}{2}}\nonumber \\
=G\left(
\begin{array}{cccc}
p_1& p_2& p_3& p_6\\
p_1& p_2& p_3& p_6 
\end{array}
\right)^{\frac{5-D}{2}} 
\int \frac{d(l_1\cdot p_6)}{\big(a_2 (l_1\cdot p_6)^2 +a_1 (l_1\cdot
  p_6)+a_0\big)^{1/2}}\nonumber \\
=G\left(
\begin{array}{cccc}
p_1& p_2& p_3& p_6\\
p_1& p_2& p_3& p_6 
\end{array}
\right)^{\frac{5-D}{2}}  \frac{1}{\sqrt{a_0}}\,.
\label{6dLS_max_cut_simplfied}
\end{gather}
Here we rationalized the square root and compute a contour integral. The
constant $a_0$ is,
\begin{gather}
  a_0 = -G\left(
\begin{array}{cc}
p_4^\sharp& p_{123}\\
p_4^\sharp& p_{123}
\end{array}
\right).
\end{gather}

Thus, by choosing the following factor 
\begin{align}
     F_4&= \frac{\epsilon_{1236}}{4}\sqrt{-p_4^{\sharp 2}s_{123}+(p_4^\sharp \cdot p_{123})^2}\\
&=\frac{1}{8}\bigg( 16 G(1,2,3,6) s_{45}^2 + 16 G(1,2,3,5) s_{46}^2 +  16 G(1,2,3,4) s_{56}^2
  \nonumber \\& - 2 \epsilon_{1235} \epsilon_{1236} s_{45} s_{46} - 2
                \epsilon_{1234} \epsilon_{1236} s_{45} s_{56} - 2
                \epsilon_{1234} \epsilon_{1235} s_{46} s_{56}
                \bigg)^{1/2} \, ,
\label{dpa_F4}
 \end{align}
we get the UT integral \eqref{dpa_UT_3}. In the second line, we
express $F_4$ with more familiar notations. 

Note that the square root in $F_4$ is {\it not} rationalized by the
momentum twistor parametrization. We will see that this square root leads
to several symbol letters which are not rationalized by momentum
twistor parametrization.

\item We find that the even combination
  \begin{equation}
    \label{eq:28}
    N_1^\text{DP-a}+N_4^\text{DP-a}
  \end{equation}
is not a UT integral by itself.
Although it has constant four-dimensional leading singularity, the Baikov
loop-by-loop analysis reveals a further, non-constant residue.
The latter can be cancelled by adding a suitable
$5\times 5$ Gram determinant to the numerator,
\begin{equation}
        N_1^\text{DP-a}+N_4^\text{DP-a} +F_5 \mu_{12} \,.
\end{equation}
Here $F_5$ is a function of kinematic variables. 
\end{itemize}

We list the analytic form of $F_3$, $F_4$ and $F_5$ in terms of
momentum twistor parametrization variables in the auxiliary file of
this paper.

We verified that, using the five integrals given in Eqs. (\ref{eqDoublePentagonBaikov1})-(\ref{eqDoublePentagonBaikov5}) master integrals, the differential equations on the cut take the canonical form (\ref{DEcanonical}).
From the analytic differential equation, the symbol letters are
identified. Many letters are either simple combinations of the
Mandelstam variables, pseudo scalars or letters for two-loop five-point
integrals with one mass \cite{Abreu:2020jxa}. The square root
$\eqref{dpa_F4}$ leads to several letters with the form, 
\begin{equation}
  \label{eq:36}
  \frac{F_4+R_i}{F_4 -R_i},\quad  i=1,\ldots,5 \, ,
\end{equation}
where each $R_i$ is a rational function in momentum twistor
parametrization variables. The explicit form of $R_i$'s are given in
the auxiliary file.

\subsection{Pentagon-Box}
The pentagon-box sector, called DP-b, is depicted in Fig.~\ref{figure:Genuine}(b).
There are three master integrals in this sector.

It is easy to construct the UT integrals with chiral numerators, similar to the previous section. We find
\begin{align}
  \label{eq:b2}
  N_1^\text{DP-b}&= s_{12} \bigg(s_{16}+\frac{\langle 26 \rangle [23][61]
  }{[13]}\bigg) s_{56} (l_1- w_1)^2\, , \\
 N_2^\text{DP-b}&= s_{12} \bigg(s_{16}+\frac{[ 26] \langle 23\rangle\langle 61 \rangle
  }{\langle 13\rangle}\bigg) s_{56} (l_1- w_2)^2\, ,\\
N_3^\text{DP-b}&=2 s_{12} s_{56} l_1\cdot (w_1-w_2) (l_1+p_6)^2\, ,
\end{align}
where the vectors $w_1$ and $w_2$ are
\begin{align}
  \label{eq:b3}
  w_1=p_1+\frac{[23]}{[13]}\lambda_2 \tilde \lambda_1,\quad w_2= w_1^* \,.
\end{align}
The integrals with these numerators have constant 4D leading
singularity, so we denote the three integrals corresponding to the above three
integrands $I_i^\text{DP-b}$, $i=1,2,3$. We comment that the
$D$-dimensional Baikov
leading singularity analysis again suggests that they are UT integral candidates.

Employing the finite-field methods for IBP reduction and rational reconstruction, we find that with this choice of basis, the cut differential equations are canonical.

Most alphabet letters appearing in them can be expressed in terms of the ones from 
known subsectors.

There is one new letter,
\begin{equation}
  \label{eq:9}
  W_{139} =s_{12} \epsilon_{1456}+s_{123} \epsilon_{1256} \,.
\end{equation}

\subsection{Pentagon-Triangle}

The pentagon-triangle integral sector DP-c is depicted in Fig.~\ref{figure:Genuine}(c).
There is only one master integral in this top sector. 

A loop-by-loop leading singularity analysis suggests to define $I_1^\text{DP-c}$ to have the following
integrand,
\begin{align}
  \label{dpc_numerator}
  N_1^\text{DP-c}= s_{13} \bigg(s_{56}-\frac{\langle 43\rangle
  \langle 15\rangle [54]}{\langle 13 \rangle }-\frac{\langle 63\rangle
  \langle 15\rangle [56]}{\langle 13 \rangle }\bigg)\bigg(l_1+\frac{Q_{456}\cdot \tilde \lambda_3 \tilde
  \lambda_1}{[13]}\bigg)^2\,.
\end{align}
We checked that the corresponding cut differential equation is
canonical. Note that the differential equation contains a special
even letter,
\begin{align}
  \label{eq:23}
  W_{182}&=\bigg(s_{56}-\frac{\langle 43\rangle
  \langle 15\rangle [54]}{\langle 13 \rangle }-\frac{\langle 63\rangle
  \langle 15\rangle [56]}{\langle 13 \rangle }\bigg) \bigg(
\text{helicity  conjugate}\bigg)\nonumber \\
&=\frac{s_{15} s_{34} s_{45} + s_{15} s_{36} s_{45} -s_{15} s_{35} s_{46} + s_{14} s_{35} s_{56} +s_{16} s_{35} s_{56} -
  s_{13} s_{45} s_{56}}{s_{13}} \, ,
\end{align}
where the chiral numerator factor in \eqref{dpc_numerator} explicitly appears.

\subsection{Double Box} 
\label{subsection: double box}
Let us consider the double box integral sector DP-d, as shown in Fig.~\ref{figure:Genuine}(d).
There are $7$ master integrals in this sector. 

A four-dimensional integrand analysis yields the following numerators whose residues on the heptacut are all rational numbers (here $ l_2'\equiv l_2-p_6$):
\begin{align}
 \label{dpd-numerators}
  N_1^\text{DP-d}&= s_{12} s_{45}s_{156}\, ,\\
 N_2^\text{DP-d}&= s_{12} s_{45} (l_1+p_5+p_6)^2\, ,\\
N_3^\text{DP-d}&=s_{12} s_{45} (l_2'+p_1+p_6)^2\, ,\\
 N_4^\text{DP-d}&=s_{45} \frac{\langle 24\rangle}{\langle 14\rangle} (\langle 15\rangle [52]+\langle 16\rangle [62]) (l_1-p_1-v_1)^2\, ,\\
  N_5^\text{DP-d}&=s_{45}\frac{[24]}{[14]}([15]\langle 52\rangle+[16]\langle 62\rangle) (l_1-p_1-v_2)^2\, ,\\
   N_6^\text{DP-d}&=s_{12}\frac{\langle 42\rangle}{\langle 52\rangle}(\langle 51\rangle[14]+\langle 56\rangle[64]) (l_2'-p_5-u_1)^2\,,\\
N_7^\text{DP-d}&=s_{12}\frac{[42]}{[52]}([51]\langle 14\rangle+[56]\langle 64\rangle) (l_2'-p_5-u_2)^2\,,\\
N_{8}^\text{DP-d}&=s_{24}\frac{\langle42\rangle\langle15\rangle}{\langle14\rangle\langle52\rangle}(l_1-p_1-v_1)^2(l_2'-p_5-u_1)^2\,,\\
N_{9}^\text{DP-d}&=s_{24}\frac{[42][15]}{[14][52]}(l_1-p_1-v_2)^2(l_2'-p_5-u_2)^2\,\\
N_{10}^\text{DP-d}&=s_{24}(l_1-p_1-v_2)^2(l_2'-p_5-u_1)^2\, ,\\
N_{11}^\text{DP-d}&=s_{24}(l_1-p_1-v_1)^2(l_2'-p_5-u_2)^2\,,
\end{align}
where the vectors $v_1$ ,$v_2$ and $u_1$,$u_2$ are
\begin{align}
  \label{eq:6}
  v_1&=\frac{\langle14\rangle}{\langle24\rangle}{ \lambda_2 \tilde
  \lambda_1},\quad v_2= v_1^* \, ,\\
  u_1&=\frac{\langle52\rangle}{\langle42\rangle}{\lambda_4 \tilde
  \lambda_5},\quad u_2= u_1^* \,.
\end{align}

However, similar to the case of double pentagon discussed above, some of these expressions need to be modified to give UT integrals.
We use the $D$-dimensional
Baikov analysis~\cite{Baikov:1996cd}, to find the following UT
integrals on the cut of this sector:
\begin{align}
  \label{eq:8}
   I_1^\text{DP-d} & =\int \frac{d^{4-2\eps} l_1}{i \pi^{2-\eps}} \frac{d^{4-2\eps} l_2}{i
  \pi^{2-\eps}} \frac{N_1^\text{DP-d}}{D_1 D_2 D_3 D_5 D_7 D_8 D_9} \, \\
  I_2^\text{DP-d} & =\int \frac{d^{4-2\eps} l_1}{i \pi^{2-\eps}} \frac{d^{4-2\eps} l_2}{i
  \pi^{2-\eps}} \frac{N_4^\text{DP-d}+N_5^\text{DP-d}}{D_1 D_2 D_3 D_5 D_7 D_8 D_9}\, ,\\
 I_3^\text{DP-d} & =\int \frac{d^{4-2\eps} l_1}{i \pi^{2-\eps}} \frac{d^{4-2\eps} l_2}{i
  \pi^{2-\eps}} \frac{N_5^\text{DP-d}}{D_1 D_2 D_3 D_5 D_7 D_8 D_9}\, ,\\
I_4^\text{DP-d} & =\int \frac{d^{4-2\eps} l_1}{i \pi^{2-\eps}} \frac{d^{4-2\eps} l_2}{i
  \pi^{2-\eps}} \frac{N_6^\text{DP-d}}{D_1 D_2 D_3 D_5 D_7 D_8 D_9}\, ,\\
I_5^\text{DP-d} &= H_1\int \frac{d^{4-2\eps} l_1}{i \pi^{2-\eps}} \frac{d^{4-2\eps} l_2}{i
  \pi^{2-\eps}} \frac{G\left(
                   \begin{array}{ccccc}
l_1& p_1& p_2& p_5& p_6\\
l_2& p_1& p_2& p_5& p_6
                   \end{array}\right)}{D_1 D_2 D_3 D_5 D_7 D_8 D_9}\, ,\label{dpd_UT_5}\\
I_6^\text{DP-d} & =\int \frac{d^{4-2\eps} l_1}{i \pi^{2-\eps}} \frac{d^{4-2\eps} l_2}{i
  \pi^{2-\eps}} \frac{N_8^\text{DP-d} +H_2 G\left(
                   \begin{array}{ccccc}
l_1& p_1& p_2& p_5& p_6\\
l_2& p_1& p_2& p_5& p_6
                   \end{array}\right)}{D_1 D_2 D_3 D_5 D_7 D_8 D_9}\, ,\label{dpd_UT_6} 
\\
I_7^\text{DP-d} &=  H_3 I_2^\text{DP-d}  + H_4 I_3^\text{DP-d} +H_5
                  I_5^\text{DP-d} 
\nonumber \\
&+\frac{2\epsilon-1}{\epsilon} H_6 \int \frac{d^{4-2\eps} l_1}{i \pi^{2-\eps}} \frac{d^{4-2\eps} l_2}{i
  \pi^{2-\eps}} \frac{G\left(
                   \begin{array}{ccccc}
l_1& p_1& p_2& p_5& p_6\\
l_1& p_1& p_2& p_5& p_6
                   \end{array}\right)}{D_1 D_2 D_3 D_5 D_7 D_8 D_9}  \,.
\label{dpd_UT_7}
\end{align}
Here $H_1,\ldots H_6$ are functions in kinematic variables which are
obtained in the Baikov leading singularity computation. They are rational functions in the momentum twistor parametrization (\ref{momentum_parametrization}). 
Some of them have simple expression in traditional kinematic variables, for example,
\begin{align}
  \label{eq:34}
  H_1&=-\frac{1}{2} \frac{\epsilon_{1245}}{G(1,2,5,6)}\,,\\
  H_6&=\frac{s_{45} \epsilon_{1234}}{\epsilon_{1256}\big\lbrack \epsilon_{2345} (s_{15}+s_{16})+\epsilon_{1345} (s_{23}+s_{24})- \epsilon_{1245} s_{34}\big\rbrack}.
\end{align}
The expressions of $H_i$'s are listed in the auxiliary file of this paper.

We proceeded as follows to obtain these expressions.
Like the double pentagon case, \eqref{dpd_UT_5} is constructed from
a $5\times 5$ Gram determinant, with the overall coefficient being obtained by a
Baikov leading singularity analysis. \eqref{dpd_UT_6} originated from $N_8^\text{DP-d}$, but includes
a $D$-dimensional correction to the numerator. Indeed, the role of the $H_2$ term is to cancel the non-constant $D$-dimensional residue for $N_8^\text{DP-d}$.
After the first $6$ UT integrals are obtained, we find that the cut differential
equation is almost canonical, except for the last row. 
We then found the final integral \label{dpd_UT_7} by a simple basis transformation.

We find that these $7$ integrals satisfy a canonical differential equation on the heptacut. 
We can read off the symbol letters appearing in this equation. 
In this block, all letters are rational in the momentum
twistor parameterization (\ref{momentum_parametrization}). 
Most letters are from five-point integrals with one massive external leg, except for the following new letter,
\begin{equation}
  \label{eq:10}
  W_{145}=\epsilon_{2345} (s_{15}+s_{16})+\epsilon_{1345} (s_{23}+s_{24})- \epsilon_{1245} s_{34}\,.
\end{equation}

\subsection{Hexagon-Box}
The hexagon-box integral sector HB-a is shown in Fig.~\ref{figure:Genuine}(e).
The only one UT integral has been shown in Section~\ref{example:hb-a}, which numerator is 
\begin{align}
  N_1^\text{HB-a}=\frac{G\left (
  \begin{array}{ccccc}
    l_1& p_1 & p_2 & p_3 & p_4 \\
 l_1& p_1 & p_2 & p_3 & p_4 
  \end{array}
\right)}{G\left(\begin{array}{cccc}
    p_1 & p_2 & p_3 & p_4 \\
 p_1 & p_2 & p_3 & p_4 
  \end{array}
\right)^{1/2}} s_{56} (l_1+p_6)^2\,.
\label{HB-a:UT}
\end{align}

We checked that the corresponding cut differential equation is in
the epsilon form. The differential
equation contains the following new symbol
letter
\begin{align}
  \label{eq:22}
  W_{176}&\equiv s_{56}+\frac{\la 12 \ra [23] \la 34 \ra [45] \la 56 \ra [61]-\la 23 \ra [34] \la
45 \ra [56] \la 61 \ra [12]}{\epsilon_{1234}} \,.
\end{align}
Interestingly, it can be expressed as $W_{176}=2(l_1^* \cdot p_5)$,
where $l_1^*$ is one of maximal cut solutions for $l_1$.

\subsection{Hexagon-Bubble}

The sector $(1,1,1,1,1,0,1,0,0,1,0)$ of this family is
named as HB-b (Figure~\ref{figure:Genuine}(f)). Based on the cut IBP compuation,
there is one master
integral in this sector. 

By the loop-by-loop Baikov analysis and the conventional bubble loop
treatment, we defined the following numerator for the UT integral in
this sector:
\begin{align}
  \label{eq:4}
  N_1^\text{HB-b}=\frac{1}{\epsilon}\frac{G\left (
  \begin{array}{ccccc}
    l_1& p_1 & p_2 & p_3 & p_4 \\
 l_1& p_1 & p_2 & p_3 & p_4 
  \end{array}
\right)}{G\left(\begin{array}{cccc}
    p_1 & p_2 & p_3 & p_4 \\
 p_1 & p_2 & p_3 & p_4 
  \end{array}
\right)^{1/2}} \frac{(l_1+p_6)^2}{(l_2-p_6)^2}\, ,
\end{align}
where again the ratio between two Gram determinants cancelled the
residue in the left hexagon loop, and the denominator $(l_2-p_6)^2$ just servers as a double propagator.

We explicitly checked that the corresponding differential equation on-cut is in
the epsilon form.

\section{Result for function alphabets}
In this section, we list the following alphabets, in the analytic form, for the canonical differential
equations, on the sector cut level.

The letters include those known from the two-loop five-point integrals with one
  off-shell leg \cite{Abreu:2020jxa}, suitably rewritten in our six-particle kinematics, including cyclic permutations.
  We denote $\rm T$ as the generator of the cyclic permutation
group of the external legs,
\begin{equation}
  \label{eq:14}
  {\rm{T}} (p_i) \equiv p_{i+1}\,,\quad i=1,\ldots 6 \,.
\end{equation}
For the new letters we identify, we also include their cyclic permutations.

With the permutations included, we can make a connection to the dual conformal hexagon function alphabet known from ${\mathcal N}=4$ super Yang-Mills.
We show how the latter is expressed in terms of our alphabet. 

We classify the alphabet as three groups, according to the following transformations (1) the
space parity (2) the sign change of the square root in the function
$F_4$ in \eqref{dpa_F4}. 

Finally, for convenience of the reader, we list separately the letters that appear in each of the integral sectors considered in this paper. Moreover, the alphabet and the cut differential equations in terms of these letters are given in the auxiliary files of this paper.

\subsection{Even letters}
\label{subsection:even_letters}
We call a letter {\it even}, if this letter's dlog is invariant under both the space parity
transformation and the sign change of $F_4$. In this subsection, we
list the following even letters:

Define
\begin{gather}
  \label{eq:12}
  W_1 = s_{12}, \quad W_{i+1}={\rm{T}}^i W_1,\quad  i=1,\ldots 5\, ,\\
W_7 = s_{13}, \quad W_{i+7}={\rm{T}}^i W_7,\quad  i=1,\ldots 5\, ,\\
W_{13}=s_{14}, \quad W_{i+13}={\rm{T}}^i W_{13},\quad  i=1,\ldots 2\, ,\\
W_{16}=s_{12} + s_{13}, \quad W_{i+16}={\rm{T}}^i W_{16},\quad
i=1,\ldots 5 \, ,\\
W_{22}=s_{12} + s_{23}, \quad W_{i+22}={\rm{T}}^i W_{22},\quad
i=1,\ldots 5 \, ,\\
W_{28}=s_{13} + s_{23},\quad W_{i+28}={\rm{T}}^i W_{28},\quad
i=1,\ldots 5 \, ,\\
W_{34}=s_{14}+s_{24} ,\quad W_{i+34}={\rm{T}}^i W_{34},\quad
i=1,\ldots 5 \, ,\\
W_{40}=s_{13}+s_{14} ,\quad W_{i+40}={\rm{T}}^i W_{40},\quad
i=1,\ldots 5 \, ,\\
W_{46}=s_{456},\quad W_{i+46}={\rm{T}}^i W_{46},\quad
i=1,\ldots 2 \, ,\\
W_{49}=s_{124},\quad W_{i+49}={\rm{T}}^i W_{49},\quad
i=1,\ldots 5 \, ,\\
W_{55}= -s_{15} - s_{16} + s_{45} + s_{46} \quad W_{i+55}={\rm{T}}^i W_{55},\quad
i=1,\ldots 2 \, ,\\
W_{58}=-s_{14} + s_{23} + s_{56}\quad W_{i+58}={\rm{T}}^i W_{58},\quad
i=1,\ldots 2 \, ,\\
W_{61}=-s_{14} - s_{23} + s_{56}\quad W_{i+61}={\rm{T}}^i W_{61},\quad
i=1,\ldots 5 \, .
\end{gather}
There are $66$ letters which are linear in $s_{ij}$'s. Here $\rm
T$ means the generator of the cyclic permutation. Note that
$W_{13}$, $W_{46}$, $W_{55}$ and $W_{58}$ are invariant under the
permutation,
\begin{equation}
  \label{eq:37}
  \rm T^3: \quad 1 \leftrightarrow 4, \quad 2 \leftrightarrow 5, \quad 3
  \leftrightarrow 6 \, ,
\end{equation}
which is the generator of the
$Z_2$ subgroup of the cyclic permutation group.

There are $57$ letters
quadratic in $s_{ij}$'s ($W_{67}\sim W_{123}$): 
\begin{gather}
  \label{eq:13}
  W_{67}=-(s_{12}s_{45}) + s_{34}s_{45} - s_{12}s_{46} + s_{34}s_{46} +
  s_{34}s_{56},\quad W_{i+67}={\rm{T}}^i W_{67},\, i=1,\ldots,5 \, , \quad\\
 W_{73}=s_{12}s_{15} +
  s_{12}s_{16} - s_{15}s_{34} - s_{16}s_{34} + s_{12} s_{56}, \quad W_{i+73}={\rm{T}}^i W_{73},\, i=1,\ldots,5 \, , \\
W_{79}=-s_{15}s_{45} - s_{16}s_{45} -
  s_{15}s_{46} - s_{16}s_{46} - s_{15}s_{56} - s_{16}s_{56} + s_{23}s_{56} - s_{45}s_{56} - s_{46}s_{56}
  - s_{56}^2,\nonumber \\
W_{i+79}={\rm{T}}^i W_{79},\, i=1,\ldots,2 \, , \\
W_{82}= s_{12}s_{15} + s_{15}^2 + s_{12}s_{16} + 2s_{15}s_{16} + s_{16}^2 - s_{15}s_{34} - s_{16}s_{34} +
 s_{12}s_{56} + s_{15}s_{56} + s_{16}s_{56},\nonumber \\
W_{i+82}={\rm{T}}^i W_{82},\, i=1,\ldots,5 \, ,
\end{gather}
\begin{gather}
W_{88}= -s_{15}s_{45} - s_{16}s_{45} - s_{15}s_{46} - s_{16}s_{46}
 + s_{12}s_{56} - s_{15}s_{56} - s_{16}s_{56} - s_{45}s_{56} -
 s_{46}s_{56} - s_{56}^2, \nonumber\\
W_{i+88}={\rm{T}}^i W_{88},\, i=1,\ldots,5 \, , \\
W_{94}=-s_{15}s_{45}
 - s_{16}s_{45} - s_{15}s_{46} - s_{16}s_{46} - s_{15}s_{56} - s_{16}s_{56} + s_{34}s_{56} - s_{45}s_{56}
 - s_{46}s_{56} - s_{56}^2,\nonumber \\
W_{i+94}={\rm{T}}^i W_{94},\, i=1,\ldots,5 \, , \\
 W_{100}=s_{15}s_{45} + s_{16}s_{45} - s_{34}s_{45} + s_{15}s_{46} +
 s_{16}s_{46} - s_{34}s_{46} + s_{15}s_{56} \nonumber \\+
 s_{16}s_{56} - s_{23}s_{56} - s_{34}s_{56} + s_{45}s_{56} +
 s_{46}s_{56} + s_{56}^2, \nonumber \\
W_{i+100}={\rm{T}}^i W_{100},\, i=1,\ldots,5\, , \\
W_{106}=-s_{12}s_{15} -
 s_{12}s_{16} + s_{15}s_{45} + s_{16}s_{45} + s_{15}s_{46} + s_{16}s_{46}  \nonumber \\- s_{12}s_{56} + s_{15}s_{56} +
 s_{16}s_{56} - s_{23}s_{56} + s_{45}s_{56} + s_{46}s_{56} + s_{56}^2,
 \nonumber\\
W_{i+106}={\rm{T}}^i W_{106},\, i=1,\ldots,5 \, ,\\
W_{112}=s_{12}s_{45} + s_{15}s_{45} + s_{16}s_{45} - s_{34}s_{45} + s_{12}s_{46} + s_{15}s_{46} + s_{16}s_{46} -
s_{34}s_{46} + \nonumber \\ s_{15}s_{56} + s_{16}s_{56} - s_{23}s_{56} - s_{34}s_{56} + s_{45}s_{56} + s_{46}s_{56} +
s_{56}^2, \nonumber \\
W_{i+112}={\rm{T}}^i W_{112},\, i=1,\ldots,5 \, , \\
W_{118}=-s_{12}s_{15} - s_{12}s_{16} + s_{15}s_{34} + s_{16}s_{34} +
s_{15}s_{45} + s_{16}s_{45} + s_{15}s_{46} + \nonumber \\ s_{16}s_{46}
- s_{12}s_{56} + s_{15}s_{56} + s_{16}s_{56} - s_{23}s_{56} +
s_{45}s_{56} + s_{46}s_{56} + s_{56}^2\nonumber \\
W_{i+118}={\rm{T}}^i W_{118},\, i=1,\ldots,5 \, .
\end{gather}
Note that $W_{79}$ is invariant under $\rm T^3$. 

The following $15$ letters ($W_{124}\sim W_{138}$) are pseudo scalars:
\begin{gather}
W_{124}=\eps_{1234},\quad W_{i+124}={\rm{T}}^i W_{124},\quad
i=1,\ldots 5 \, ,\nonumber \\
W_{130}=\eps_{1235},\quad W_{i+130}={\rm{T}}^i W_{130},\quad
i=1,\ldots 5 \, ,\nonumber \\
W_{136}=\eps_{1245},\quad W_{i+136}={\rm{T}}^i W_{136},\quad
i=1,\ldots 2 \, .
\end{gather}
Note that $W_{136}$  is invariant under $\rm T^3$. Under the
parity transformation, a pesudo
scalar transforms as $\epsilon_{ijkl} \to -\epsilon_{ijkl}$ but
$d\log(\epsilon_{ijkl})$ is invariant.

The rest even letters  ($W_{139}\sim W_{145}$, $W_{176}\sim W_{190}$)  are found through the analytic canonical
differential equations on the maximal cut:
\begin{gather}
  \label{eq:15}
  W_{139}=s_{12}\eps_{1456}+s_{123} \epsilon_{1256},\quad W_{i+139}={\rm{T}}^i W_{139},\quad
i=1,\ldots 5 \, , \\
W_{145}
=\la 12 \ra [23] \la 34 \ra [45] \la 56 \ra [61]-\la 23 \ra [34] \la
45 \ra [56] \la 61 \ra [12] \\
W_{176}=s_{56}+\frac{W_{145}}{\epsilon_{1234}},\quad W_{i+176}={\rm{T}}^i W_{176},\quad
i=1,\ldots 5  \, , \\
\begin{aligned}
W_{182}
  =\frac{s_{15} s_{34} s_{45} + s_{15} s_{36} s_{45} -s_{15} s_{35} s_{46} + s_{14} s_{35} s_{56} +s_{16} s_{35} s_{56} -
  s_{13} s_{45} s_{56}}{s_{13}}\\
  \end{aligned}
\\
W_{i+182}={\rm{T}}^i W_{182},\quad
i=1,\ldots 5 \, ,\\
W_{188}=\frac{1}{8}\bigg( 16 G(1,2,3,6) s_{45}^2 + 16 G(1,2,3,5) s_{46}^2 +  16 G(1,2,3,4) s_{56}^2
  \nonumber \\ - 2 \epsilon_{1235} \epsilon_{1236} s_{45} s_{46} - 2
  \epsilon_{1234} \epsilon_{1236} s_{45} s_{56} - 2 \epsilon_{1234}
  \epsilon_{1235} s_{46} s_{56} \bigg)^{1/2}\\
W_{i+188}={\rm{T}}^i W_{188},\quad
i=1,\ldots 2 \, .
\end{gather}
Note that under the parity transform, $W_{145}\to -W_{145}$ and $d\log
W_{145}$ is invariant. Furthermore $W_{145}$ has the remarkable property that $d\log W_{145}$  is cyclically
invariant. $W_{188}$, which equals the square root $F_4$ \eqref{dpa_F4} from the six-dimensional double pentagon UT integral, is invariant
under $\rm{T}^3$.

The dlog of the letters in this subsection are all invariant under the
parity transformation. Furthermore the dlog of these letters are invariant under the sign change of
the square root in $F_4$ \eqref{dpa_F4}. Thus we define them as even letters. The above even letters are closed under the cyclic group
permutations. The group action is manifest in the definitions
above.

\subsection{Parity odd letters}
Here a parity odd letter refers to a letter $W$ such that
$d\log W\to -d\log W$ under the space parity transformation, while
$d\log W\to d\log W$ under the sign change of $F_4$.

We use the following $30$ partity odd letters ($W_{146}\sim W_{175}$):
\begin{gather}
  \label{eq:16}
  W_{146}=\frac{\Tr_+(1245) + \Tr_+(1246)}{\Tr_-(1245) + \Tr_-(1246)}, \quad W_{i+146}={\rm{T}}^i W_{146},\quad
i=1,\ldots 5 \, ,\\
W_{152}=\frac{\Tr_+(1345) + \Tr_+(1346)}{\Tr_-(1345) + \Tr_-(1346)}, \quad W_{i+152}={\rm{T}}^i W_{152},\quad
i=1,\ldots 2 \, ,\\
W_{155}=\frac{\Tr_+(1234) }{ \Tr_-(1234)}, \quad W_{i+155}={\rm{T}}^i W_{155},\quad
i=1,\ldots 2 \, ,\\
W_{158}=\frac{\Tr_+(1235) + \Tr_+(1236)}{\Tr_-(1235) + \Tr_-(1236)}, \quad W_{i+158}={\rm{T}}^i W_{158},\quad
i=1,\ldots 5 \, ,\\
W_{164}=\frac{\Tr_+(2345) + \Tr_+(2346)}{\Tr_-(2345) + \Tr_-(2346)}, \quad W_{i+164}={\rm{T}}^i W_{164},\quad
i=1,\ldots 5 \, ,\\
W_{170}=\frac{F+\epsilon_{1234}}{F-\epsilon_{1234}},\quad W_{i+170}={\rm{T}}^i W_{170},\quad
i=1,\ldots 5 \, .
\end{gather}
Here $F$ is the polynomial,
\begin{gather}
  \label{eq:17}
  F=s_{12} s_{15} + s_{12} s_{16} - s_{12} s_{23} + s_{23} s_{34} - s_{15} s_{45} - s_{16} s_{45} + 
 2 s_{23} s_{45} + s_{34} s_{45} - s_{15} s_{46} \nonumber \\- s_{16} s_{46} + 2 s_{23} s_{46} + s_{34} s_{46} + 
 s_{12} s_{56} - s_{15} s_{56} - s_{16} s_{56} + s_{23} s_{56} + s_{34} s_{56} - s_{45} s_{56} - 
 s_{46} s_{56} - s_{56}^2\,.
\end{gather}
All the above $30$ odd letters are from the $2$-loop $5$-point one
massive alphabet permutations. Again these $30$ odd letters are closed in the cyclic
permutation.

\subsection{Letters associated with the double pentagon square root}
The following $15$ letters $(W_{191}\sim W_{205})$ are from the double pentagon cut differential
equations, 
\begin{gather}
  \label{eq:31}
  W_{191} =\frac{F_4 - R_1}{F_4+R_1}, \quad W_{i+191}={\rm{T}}^i W_{191},\quad
i=1,\ldots 2 \, ,\\
W_{194} =\frac{F_4 - R_2}{F_4+R_2}, \quad W_{i+194}={\rm{T}}^i W_{194},\quad
i=1,\ldots 2 \, ,\\
W_{197} =\frac{F_4 - R_3}{F_4+R_3}, \quad W_{i+197}={\rm{T}}^i W_{197},\quad
i=1,\ldots 2 \, ,\\
W_{200} =\frac{F_4 - R_4}{F_4+R_4}, \quad W_{i+200}={\rm{T}}^i W_{200},\quad
i=1,\ldots 2 \, ,\\
W_{203} =\frac{F_4 - R_5}{F_4+R_5}, \quad W_{i+203}={\rm{T}}^i W_{203},\quad
i=1,\ldots 2\, .
\end{gather}
Here the function $F_4$ is defined in \eqref{dpa_F4}. It contains a square root
not rationalized with the momentum twistor parameterization. $R_i$,
$i=1,\ldots, 5$ are rational functions in the momentum twistor
parameterization variables. The expression for $R_i$'s are given in
the auxiliary file of this paper. 

These $15$ letters are
closed under the cyclic permutations. Under the sign change of the square root in $F_4$, all these letters
transform as $d \log W\to -d\log W$. Under the space
parity transformation, however, these letters transform as,
\begin{gather}
  \label{eq:38}
  d\log W_{191} \to -d\log W_{194},\quad d\log W_{194} \to -d\log
  W_{191},
\\
d\log W_{197} \to -d\log W_{191}+d\log W_{194}-d\log W_{197}\\
d\log W_{200} \to -d\log W_{200},\quad d\log W_{203} \to -d\log
  W_{203},
\end{gather}
It is possible to diagonalize the parity transform by
recombining these letters, however,
the resulting letters would have longer expressions.

 \subsection{Relation to dual conformal hexagon function alphabet}

From the study of hexagon remainder
function \cite{Dixon:2011pw,Dixon:2013eka}, the following variables are
well known:
\begin{equation}
  \label{eq:19}
  u_1=\frac{s_{12}s_{45}}{s_{123}s_{345}},\quad
  u_2=\frac{s_{23}s_{56}}{s_{234}s_{456}},\quad 
  u_3=\frac{s_{34}s_{16}}{s_{345}s_{156}},
\end{equation}
and $\Delta=(1-u_1-u_2-u_3)^2-4 u_1 u_2 u_3$,
\begin{equation}
  \label{eq:20}
  z_{\pm}=\frac{-1+u_1+u_2+u_3\pm\sqrt{\Delta}}{2}\,,
\end{equation}
\begin{equation}
  \label{eq:21}
  y_i=\frac{u_i-z_+}{u_i-z_-},\quad i=1,2,3\,.
\end{equation}
The hexagon remainder function alphabet \cite{Dixon:2011pw, Dixon:2013eka} is
\begin{equation}
  \label{hexagon_alphabet}
  u_1,\ u_2,\ u_3,\ 1-u_1,\ 1-u_2,\ 1-u_3,\ y_1,\ y_2,\ y_3
\end{equation}
In our momentum twistor parametrization, $\sqrt{\Delta}$ is rationalized.

Here we remark that the hexagon remainder function
  alphabet \eqref{hexagon_alphabet} are contained in the $205$ letters chosen above. Indeed, the nine hexagon alphabet letters (\ref{hexagon_alphabet}) can be written as rational functions of $W_1,\dots, W_{145}$,
\begin{gather}
u_1=\frac{W_1 W_4}{W_{46} W_{48}},\quad
  u_2=\frac{W_2 W_5}{W_{46} W_{47}},\quad 
  u_3=\frac{W_3 W_6}{W_{47} W_{48}}\\\nonumber
1-u_1=-\frac{W_{81}}{W_{46} W_{48}}, \quad 1-u_2=-\frac{W_{79}}{W_{46} W_{47}},\quad 1-u_3=-\frac{W_{80}}{W_{47} W_{48}}, \\\nonumber
y_1=W_{146} W_{153}, \quad y_2=\frac{1}{W_{147} W_{154}},\quad  y_3=\frac{1}{W_{151} W_{152}}.
\end{gather}
What is more, we note that $u_1$, $u_2$, $u_3$ and $1-u_1$, $1-u_2$, $1-u_3$ are
products of even letters, while $y_1$, $y_2$, $y_3$ are products of odd
letters.

 We also note that $\sqrt{\Delta}$ can be written as rational functions of $W_1,\dots, W_{145}$, as follows,
\begin{gather}
\sqrt{\Delta} = \frac{W_{145}}{W_{46} W_{47} W_{48}} \,.
\end{gather}

\subsection{Maximal cuts of the differential equation and alphabet letters}
In this subsection, we list the letters used for each cut differential
equation discussed in this paper. Note that the letters we listed in
this section is closed under the cyclic permutation, however for a
particular sector (with one particular orientation of external legs), only a small subset of
the letters appear in the cut differential equation.
\begin{itemize}
\item Dp-a:
\begin{gather}
W_1,W_2,W_3,W_4,W_5,W_6,W_7,W_{10},W_{46},W_{79},W_{81},W_{124},W_{126},W_{127},W_{129},
W_{145},\nonumber
\\W_{146},W_{147},W_{148},W_{151},W_{152},W_{153},W_{154},W_{188},W_{191},W_{194},W_{197},W_{200},W_{203}
\nonumber 
\end{gather}
\item DP-b:  $W_1,W_2,W_5,W_6,W_7,W_{11},W_{81},W_{120},W_{129},W_{130},W_{139},W_{149},W_{151},W_{152},W_{160}$
\item Dp-c: $W_1,W_2,W_7,W_{130},W_{145},W_{182}$
\item Dp-d: 
  \begin{gather}
    \label{eq:32}
    W_1,W_2,W_3,W_4,W_5,W_6,W_8,W_{11},W_{13},W_{14},W_{47},W_{79},W_{80},W_{82},W_{85},W_{124},W_{125},W_{127},\nonumber\\
W_{128},W_{136},W_{145},W_{146},W_{147},W_{149},W_{150},W_{151},W_{152},W_{153},W_{154},W_{155},W_{156},W_{171},W_{174}\nonumber 
  \end{gather}
\item Hb-a: $W_1,W_2,W_3,W_5,W_{79},W_{124},W_{145},W_{176}$
\item HB-b: $W_1,W_2,W_3,W_{79},W_{124},W_{145}$
\end{itemize}
The cut differential equations in terms of the $d\log$ of the symbol
letters are given in the auxiliary file of this paper.

\section{Description of ancillary files}

Here we briefly describe the content of the auxiliary files attached with
this paper. The auxililary files are included in a folder {\tt output}.
\begin{itemize}
\item In the file {\tt momentum-twistor-rep.txt}, kinematic variables like $s_{ij}$, $\epsilon_{ijkl}$ and
  $\text{Tr}_\pm(i,j,k,l)$ are expressed in terms of our momentum twistor
  parametrization in the subsection \ref{sec:kinematics}.
\item The definitions of the functions $F_i$'s, $H_i$'s in the
  subsection \ref{subsection:double pentagon} and \ref{subsection: double box}, in terms of the momentum twistor
  parametrization, are given in 
  the {\tt .txt} files with the corresponding names. The functions $R_i$,
  $i=1,\ldots 5$, are defined in the file {\tt RFactors.txt}.
\item The list of alphabet letters, are firstly formally defined in terms of $s_{ij}$,
  $s_{ijk}$, $\eps_{ijkl}$, $F_i$, $H_i$ and $R_i$, in the file {\tt
    alphabet-formal-definition.txt}. They are secondly  defined in terms of
  the momentum twistor parametrization variables, in the file {\tt alphabet.txt}.
\item  Differential equation matrices for the integral sectors considered in
  this paper, are given in the files {\tt
    `sector-name-abbreviation'-DE.txt}. Note that the DE matrices are
  given in the $d\log$ alphabet notation. 
\end{itemize}

\section{Summary and outlook}

In this paper, we studied planar two-loop six-particle Feynman integrals. We performed D-dimensional integrand analysis in order to find UT integrals on the cuts. We showed that the basis integrals found by our analysis satisfy canonical differential equations on the cut, and we identified novel alphabet letters appearing in the differential equations. This constitutes an important step towards the analytic calculation of two-loop six-particle Feynman integrals, and gives insights into their function space.

There are a number of natural directions for future work:

1. Thanks to the results in this paper, obtaining the canonical form of the differential equations for all planar, two-loop six particle Feynman integrals should be within reach. The challenge would be to obtain and simplify the off-diagonal blocks of the differential equations, namely the ones that couple the genuine six-particle sectors discussed here, to the five-point sectors with one off-shell leg. We do not expect major conceptual difficulties, however given the number of master integrals involved, dedicated finite-field methods might be needed. 

2. A second interesting direction is to establish the full symbol alphabet for two-loop planar six-particle scattering processes. 
This is extremely interesting theoretical information, and in addition could help with rational reconstruction of entries of the differential equations. 
In the present paper, we already identified a number of genuine six-particle symbol letters. Based on experience with five-particle integrals one may expect additional symbol letters in the off-diagonal blocks of the differential equations, which are yet to be computed. However, one could try to obtain these letters in an alternative, possibly simpler way. The procedure is in two steps. In a first step, one could use Landau equations, see e.g. \cite{LANDAU1959181,collins_2011,Mizera:2021icv}, to establish the allowed set of singularities. Then, in a second step, given the known square root factors (in part identified in the present paper), one can systematically construct alphabet letters that are compatible with the set of singularities. See e.g. \cite{Henn:2018cdp,Zoia:2021zmb} for a discussion.  This could constitute a useful shortcut to obtaining the symbol alphabet for these processes. 

3. Once the symbol alphabet is known, either via the differential equations, or via the Landau and algebraic analysis mentioned in the last paragraph, a third interesting connection would be to analyze further the analytic properties of the function space. 
For example, it would be very useful to classify which iterated integrals are permitted by general QFT principles, such as the branch cut structure, Steinmann relations. This would on the one hand inform bootstrap approaches (see e.g. \cite{Henn:2018cdp,Elvang:2020lue}), and on the other hand be an important step for developing numerical codes for evaluating the special functions, as e.g. in \cite{Chicherin:2020oor,Chicherin:2021dyp}.

4. In ${\mathcal{N}}=4$ super Yang-Mills, remarkably, the function space of planar six-particle amplitudes is governed by a cluster algebra \cite{Golden:2013xva,Chicherin:2020umh}. It is an interesting question whether cluster algebra structures also play a role in general quantum field theory. Knowing the full six-particle alphabet will make it possible to investigate this question further in a particularly interesting case.

\section*{Acknowledgement}
We thank Dmitry Chicherin, Christoph Dlapa, Song He, Gudrun Heinrich,
Zhenjie Li, Leila
Maestri, Pascal Wasser, Kai Yan, Peter Uwer, Qinglin Yang and Simon
Zoia for enlightening discussions.  This project received funding
from the European Union’s Horizon 2020 research and innovation
programmes New level of theoretical precision for LHC Run 2 and beyond
(grant agreement No 683211), High precision multi-jet dynamics at the
LHC (grant agreement No 772009), and Novel structures in scattering
amplitudes (grant agreement No 725110). YZ is supported from the NSF
of China through Grant No. 11947301, 12047502, 12075234 and Key Research Program of the Chinese Academy of Sciences, Grant
No. XDPB15. YZ also 
acknowledges the Institute of Theoretical Physics, Chinese Academy of Sciences, for the hospitality
through the ``Peng Huanwu visiting professor program''. Xu's research is funded by the Deutsche Forschungsgemeinschaft (DFG,
German Research Foundation) - Projektnummer 417533893/GRK2575
``Rethinking Quantum Field Theory''.

\appendix

\bibliographystyle{JHEP}
\bibliography{2l6p_Integrals}

\end{document}